\documentclass[fleqn,usenatbib]{mnras}

\usepackage{newtxtext,newtxmath}

\usepackage[T1]{fontenc}

\DeclareRobustCommand{\VAN}[3]{#2}
\let\VANthebibliography\thebibliography
\def\thebibliography{\DeclareRobustCommand{\VAN}[3]{##3}\VANthebibliography}

\usepackage{newtxtext,newtxmath}
\usepackage[T1]{fontenc}
\usepackage{ae,aecompl}
\usepackage{graphicx}	
\usepackage{amsmath}	

\newcommand{\Msun}{M\ensuremath{_\odot}}
\newcommand{\Msunyr}{M\ensuremath{_\odot}~yr\ensuremath{^{-1}}\,}

\newcommand{\Gaia}{{\it Gaia}\,}

\newcommand{\FeH}{\ensuremath{\rm [Fe/H]}\,}

\title[Why does the Milky Way have a bar?]{Why does the Milky Way have a bar?}

\author[Khoperskov et al.]{Sergey Khoperskov, Ivan Minchev, Matthias Steinmetz, Bridget Ratcliffe,   Jakob C. Walcher,  \newauthor Noam Libeskind
\\
Leibniz-Institut f\"{u}r Astrophysik Potsdam (AIP), An der Sternwarte 16, 14482 Potsdam, Germany \\
}

\date{Accepted XXX. Received YYY; in original form ZZZ}

\pubyear{2023}

\begin{document}
\label{firstpage}
\pagerange{\pageref{firstpage}--\pageref{lastpage}}
\maketitle

\begin{abstract}
There is no doubt that the Milky Way is a barred galaxy; however, factors that establish its prominent morphology remain largely elusive. In this work, we attempt to constrain the history of the MW by tracing the present-day parameters and evolution of a set of MW and M31 analogues from the TNG50 simulations. We find that the strength of bars at $z=0$ correlates well not only with the stellar mass build-up but, more crucially, with the time of onset of stellar discs. Discs of strongly barred galaxies form early ($ z \gtrsim 2-3$), compared to weakly and non-barred galaxies ($z \approx 1-1.5$). Although we are cautious to draw ultimate conclusions about the governing factor of discs formation due to the complexity and correlations between different phenomena, the observed morphological diversity can be tentatively explained by a substantial variation in the gas angular momentum around proto-galaxies already at $z\approx 3-5$; in such a way, early discs formed from gas with larger angular momentum.

By comparing the formation time scales of discs of barred galaxies in the TNG50 sample, we infer that the MW has a strong bar~($0.35<A_2<0.6$) and that its stellar disc started to dominate over the spheroidal component already at $z \approx 2$, with a mass of $\approx 1 \pm 0.5 \times 10^{10} M_\odot$. We conclude that the presence of a strong bar in the MW is a natural manifestation of the early formation of the stellar disc, which made possible bursty but highly efficient star formation at high redshift.

\end{abstract}


\maketitle

\section{Introduction}\label{sec::intro}

Since the first discovery of barred galaxies~(NGC 613, by William Herschel in 1798) and early observational studies~\citep{1918PLicO..13....9C,1926ApJ....64..321H,1959HDP....53..275D,1961hag..book.....S}, it became clear that these objects are very abundant. Recent data suggest that bars are a common feature of nearly two-thirds of disc galaxies in the local universe~\citep{2000AJ....119..536E,2000ApJ...529...93K, 2007ApJ...659.1176M,2007ApJ...657..790M, 2008ApJ...675.1141S,2011MNRAS.411.2026M}. 

In the Milky Way~(MW) context, \cite{1957AJ.....62...19J} was likely the first to raise the question of whether the MW is a barred galaxy. Nowadays, although the exact parameters of the MW bar are still debated, starting from works by \cite{1991Natur.353..140N,1991ApJ...379..631B,1994ApJ...425L..81W,1995ApJ...445..716D} we are confident that the MW is a barred galaxy. However, the question of why the MW hosts a bar still requires some understanding of mechanisms and conditions favourable to its emergence. 

Early numerical works on discs galaxy evolution demonstrated that bars naturally form due to gravitational instability~\citep[][see also a review by \cite{1975IAUS...69..297B}]{1971ApJ...168..343H,1973ApJ...186..467O} while a more detailed theory was developed in the framework of discs stability analysis~\citep{1964ApJ...139.1217T,1971ApJ...168..343H,1972ApJ...175...63K}. In particular, the bars are believed to be a superposition of leading and trailing modes inside the corotation radius, where the leading mode results from the trailing mode's reflection near the galactic centre due to swing amplification~\citep{1965MNRAS.130..125G,1966ApJ...146..810J,1981seng.proc..111T}. The latest formally requires a lack of the inner Lindblad resonance, which, however, is not a strong condition in the case of finite disc thickness in the center~\citep[`soft centre', see][]{2016MNRAS.462.3727P,2023arXiv230214775S}. Therefore, a spontaneous growth of bars seems to be a natural outcome of isolated disc evolution, which, however, can be promoted by interactions with nearby galaxies~\citep{1998ApJ...499..149M,2004MNRAS.347..220B} or suppressed by a massive central spheroid~\citep{1980A&A....89..296S,2018MNRAS.475.1653K} or due to gas inflow~\citep{2002A&A...392...83B,2007ApJ...666..189B}. Nevertheless, prominent bars are not a universal feature among disc galaxies, questioning why a substantial fraction of discs avoid bar formation~\citep{1993RPPh...56..173S, 1996ASPC...91....1F}.

An $N$-body simulation of a bar formation in isolated MW-like discs is a trivial exercise allowing to trace the physics behind the morphological transformation of discs. However, such models lack predictive power as they do not incorporate the assembly history of galaxies. In this sense, cosmological simulations are a perfect test bed for studying bar formation~\citep{2002NewA....7..155S,2007MNRAS.374.1479G,2003ApJ...591..499A, 2009MNRAS.396.1972P, 2017MNRAS.467..179G, 2018ApJ...861...88B, 2018MNRAS.473.1656T, 2019MNRAS.485.5073D}. Thanks to increasing numerical resolution and careful treatment of subgrid physics, recent large-scale cosmological simulations demonstrate the formation of bars in a variety of settings~\citep{2020ApJ...895...92Z, 2020ApJ...904..170Z, 2022MNRAS.512.5339R, 2019MNRAS.483.2721P,2017MNRAS.469.1054A,2022A&A...668L...3L,2021A&A...647A.143L}. While there is a prevailing consensus regarding the decreasing fraction of barred galaxies with increasing redshift in cosmological simulations, there remains an ongoing discussion regarding the precise quantitative comparison to observational data.

In a suite of zoom-in simulations in a cosmological context, \cite{2012ApJ...757...60K} found that the characteristic bars start to form at $z\approx 0.8-1$, at the epoch at which today's spirals acquire their disc-dominated morphology. In agreement, a study conducted by \cite{2022MNRAS.514.1006I} revealed that strongly barred galaxies in TNG50 simulations tend to have earlier assembly histories. These galaxies also have more massive and compact discs and are more dominated by stars in their central regions compared to unbarred galaxies.  Recently, using zoom-in simulations from the Auriga project~\citep{2017MNRAS.467..179G}, \cite{2021A&A...650L..16F} suggested that in order to reproduce the dynamics of barred galaxies, massive discs must be baryon-dominated and should lie above the abundance matching relation. This statement is supported by the results from large-scale cosmological HorizonAGN simulations~\citep{2022MNRAS.512..160R}, where the high fraction of dark matter explains the lack of bars in the MW-mass range disc galaxies.

\begin{figure}
\begin{center}
\includegraphics[width=1\hsize]{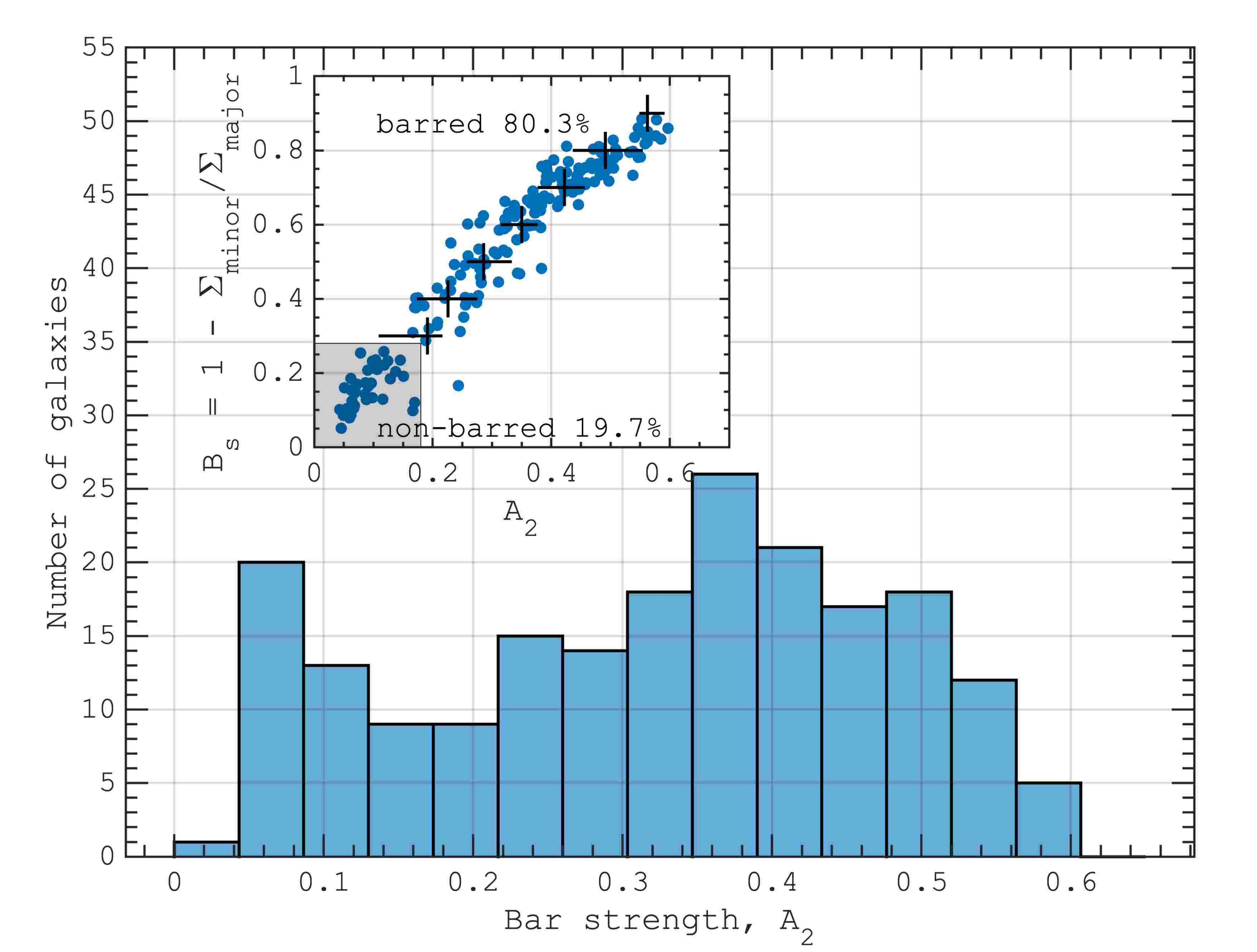}
\caption{Distribution of the bar strength at $z=0$ measured as the stellar density ratio between the major and minor axes of the bar measured inside the region with constant~($<5$ deg) density alignment. The subpanel shows a relation between the stellar density asymmetry ratio across the bar $B_s$~(see Section~\ref{sec::sample}) and the bar strength parameter $A_2$.}\label{fig::bar_strength}
\end{center}
\end{figure}

\begin{figure*}
\begin{center}
\includegraphics[width=1\hsize]{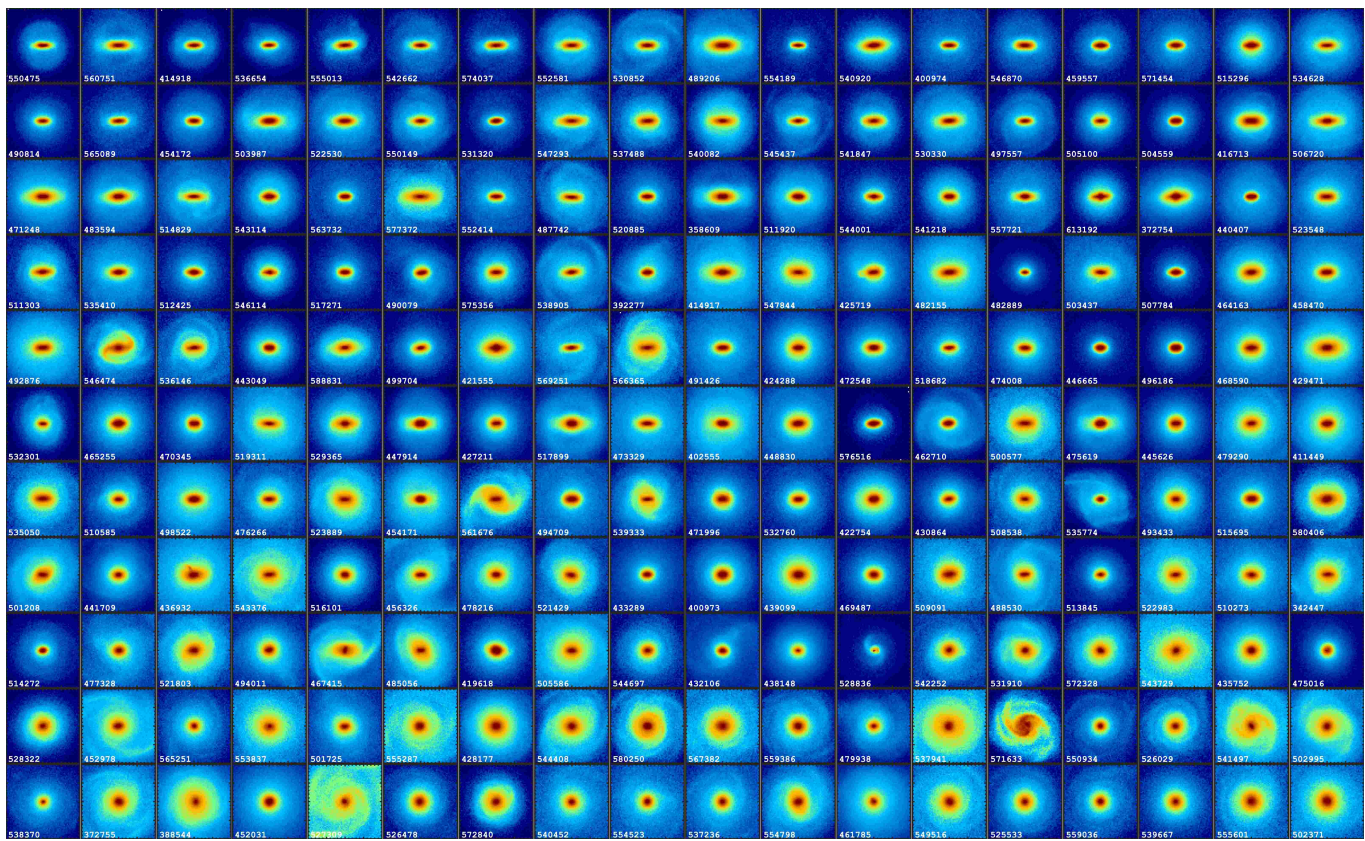}
\caption{Face-on density maps of 198 MW/M31 analogues from TNG50 simulation at $z=0$. The galaxies are sorted in descending order according to the bar strength value $A_2$. The galaxy IDs are highlighted in each panel. Each panel has a size of $\rm 15\times15~kpc^2$.}\label{fig::density_maps}
\end{center}
\end{figure*}

The aim of our study is to establish a connection between the evolution of the MW and a sample of the MW-type disc galaxies from the TNG50 cosmological simulation~\citep{2019MNRAS.490.3234N,2019MNRAS.490.3196P}. Extensive research has been conducted on the TNG50 galaxies, revealing an unparalleled degree of realism, including a thorough examination of the internal structures and galactic star formation~\citep{2019MNRAS.490.3234N, 2019MNRAS.490.3196P, 2020MNRAS.498.2391N, 2021MNRAS.501.4359Z, 2022ApJ...926..139M}. \cite{2022ApJ...940...61F} showed that bars in TNG50 simulations do not rotate slowly, which was a problem in lower-resolution preceding simulations while being shorter compared to their counterparts from MaNGA observations. \cite{2022MNRAS.512.5339R} find the bar fraction decreases with increasing redshift for TNG50 galaxies, but also barred galaxies have older stellar populations, lower gas fractions and star formation rates than unbarred galaxies, in general agreement with observational data~\citep{2012MNRAS.424.2180M, 2012MNRAS.423.3486W, 2018MNRAS.473.4731K,2020MNRAS.495.4158F}. \cite{2022MNRAS.512.2537G} find that bars in the TNG50 simulations play an important role in the development of bulges.

In this work, relying on the fact that the MW is a typical barred galaxy, we seek to understand the conditions under which the MW was able to develop the prominent bar observed in the present day. The structure of our paper is as follows. In Section~\ref{sec::sample}, we outline the sample of TNG50 galaxies used and describe the main methods employed. The analysis of the TNG sample results is presented in Section~\ref{sec::results}. In Section~\ref{sec::MW-bar} we utilize the analysis of TNG galaxies to infer the conditions favouring the formation of the MW bar. We discuss and summarize the key results of our work in Sections~\ref{sec::discus} and ~\ref{sec::concl}, respectively.

\section{MW and M31 analogues from TNG50 simulation and methods}\label{sec::sample}

In this work, we analyse a sample of the MW and M31 analogues~\citep{2023arXiv230316217P}
selected from the TNG50 simulation~\citep{2019ComAC...6....2N, 2019MNRAS.490.3234N, 2019MNRAS.490.3196P}\footnote{\url{https://www.tng-project.org/data/milkyway+andromeda/}}. TNG50 is a magneto-hydrodynamical simulation of the formation and evolution of galaxies in a 51.7 comoving Mpc cube from redshift $\approx$ 127 to redshift 0. It is run with the moving-mesh code AREPO~\citep{2010MNRAS.401..791S,2020ApJS..248...32W} and uses the fiducial TNG galaxy formation model~\citep{2017MNRAS.465.3291W,2018MNRAS.473.4077P} with a mass resolution of $m_\text{baryon} = 8.5 \times 10^4 \text{M}_\odot$, $m_\text{DM} = 4.5 \times 10^5 \text{M}_\odot$; and a spatial resolution of star-forming gas of $\sim 100-140$ pc~\citep{2019MNRAS.490.3196P}. 

The selection criteria of MW/M31 analogues at $z = 0$ include the following: (i) the galaxy stellar mass is in the following range: $\rm M_*(<30 kpc) = 10^{10.5-11.2}M_\odot$; (ii) a disc-like stellar morphology; (iii) no other galaxy with stellar mass $> 10^{10.5} M_\odot$ is within $500$~kpc distance; and (iv) the total mass of the halo host is smaller than that typical of massive groups $< 10^{13}M_\odot$. These criteria result in $198$ MW and M31 analogues, which will be our sample.

Unlike some previous works based on the TNG50 galaxies~(e.g., \citealt{2022MNRAS.512.5339R, 2022MNRAS.514.1006I}), we analyse the sample of selected galaxies as a function of the bar strength, instead of simply dividing them onto `barred' and `non-barred'. We suggest that this is a more reasonable approach because it is not evident what kind of selection criteria are physically motivated and result in a pure barred~(or non-barred) sample. This also allows us to trace a smooth transition in morphological properties of galaxies as a function of their formation paths.

Through this work, we use a traditional approach for the bar strength measurements~\citep[see][for comparison of different methods]{2020MNRAS.497..933H}: the peak value of $m=2$ Fourier harmonics of the density distribution~(i.e., the maximum value of $\rm A_2$). The value is calculated inside the inner galaxy region where the orientation of the major axis of the surface density distribution does not deviate by more than $5$ degrees over $0.5$~kpc~\citep{2002MNRAS.330...35A}. The distribution of the bar strength at $z=0$ is shown in Fig.~\ref{fig::bar_strength}, where the $\rm A_2$ values vary in the range of $0-0.6$ with a prominent bimodality. 

The asymmetry in the surface density-based definition of bar strength, specifically the $m=2$ Fourier mode, has a limitation in that it lacks direct information about the actual density structure of the bar. To establish a connection between bar strength and the 2D stellar mass distribution, in Fig.~\ref{fig::bar_strength}, we present a sub-panel that illustrates the relationship between the maximum value of $\rm A_2$~(bar strength) and the ratio of stellar densities along the major ($\rm \Sigma_{major}$) and minor ($\rm \Sigma_{minor}$) axes of the bar. The later two quantities measured from the 1D azimuthal distribution of stellar density taking into account all star particles inside the bar length. Their ratio represent the stellar density contrast across the bar and reflects the strength of the bar. We see that the traditionally measured strength of the bar, $\rm A_2$, correlates well with the density ratio~($\rm B_s = 1 - \Sigma_{minor}/\Sigma_{major}$), with a slope of $\approx 4/3$, so that a threshold of $\rm A_2=0.2$, often used for barred galaxies selection, corresponds to $\approx 30\%$ density variation across the bar. We suggest that this relation could be used to match simulated barred galaxies with observational data.

The bimodal distribution observed in the bar strength in Fig.~\ref{fig::bar_strength}, with two distinct peaks at approximately $\sim0.06$ and $\sim0.4$, indicates the presence of two populations: non-barred and barred galaxies. This suggests that there may be fundamental differences in the formation or evolution of these two populations. However, we exercise caution in drawing strong conclusions at this stage due to the incompleteness of our galaxy sample. Nonetheless, the presence of this bimodal distribution is intriguing, even if it somehow arises from the selection of MW and M31-like galaxies within the TNG50 simulation~\citep{2023arXiv230316217P}.

In order to demonstrate the difficulty of tracing a transition between 'barred' and 'non-barred' galaxies visually, we present the face-on stellar density maps of the complete dataset in Fig.~\ref{fig::density_maps}. The galaxies are arranged in descending order based on the $A_2$ parameter, and each panel highlights the corresponding IDs. This visualisation provides insight into the challenging nature of morphological identification of the transition between these two types of galaxies.

\begin{figure*}
\begin{center}
\includegraphics[width=1\hsize]{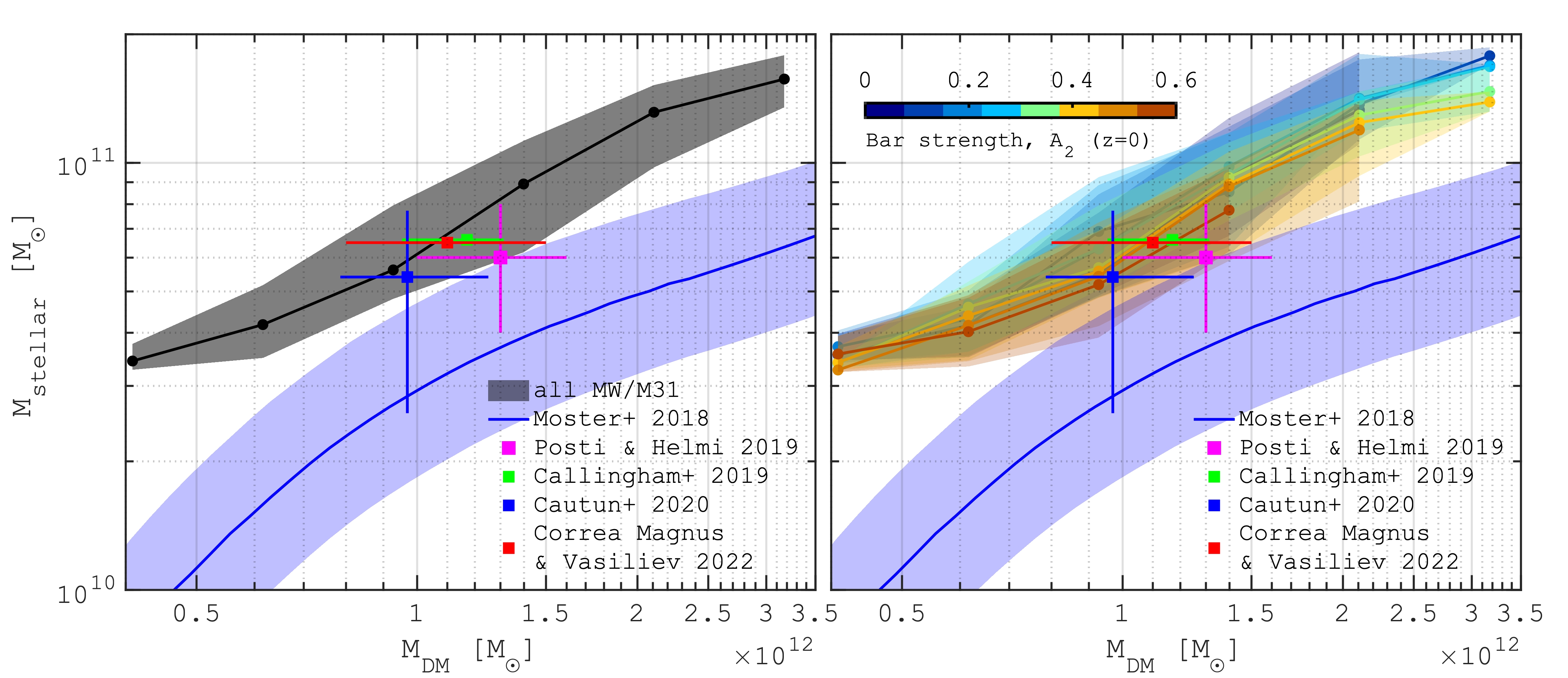}
\caption{ {\it Left:} stellar-to-dark matter mass relation for all MW and M31W TNG50 galaxies~(grey colour). {\it Right:} same but the TNG50 galaxies binned in $0.1$ range of the bar strength according to the colour bar. In both panels, the semi-analytic relation from \protect\cite{2018MNRAS.477.1822M} is shown by the blue area. Several recent estimations of the MW parameters are shown with squares of different colour with error bars~\citep{2019A&A...621A..56P,2019MNRAS.484.5453C,2020MNRAS.494.4291C,2022MNRAS.511.2610C}.}\label{fig::st2dm}
\end{center}
\end{figure*}

\section{Evolution of TNG50 galaxies with bar strength}\label{sec::results}

\subsection{Variation with baryon-to-DM ratio} 

We begin by assessing whether our sample of galaxies is dominated by baryonic matter or dark matter~(DM) at redshift zero. In Fig.~\ref{fig::st2dm}~(left), we present the relationship between the total stellar mass and DM mass for the entire sample. For reference, we include the semi-analytic relation by \cite{2018MNRAS.477.1822M} and data for the MW. 
Recent results in the literature seem to be converging on a MW mass of $M_{200, tot} \sim 10^{12} M_\odot$~\citep{2019MNRAS.487.2685E,2020MNRAS.498.5574E,2019MNRAS.485.3514D,2021MNRAS.501.5964D,2021MNRAS.501.2279V} with some providing precise estimations of baryon and dark matter masses~\citep{2019A&A...621A..56P,2019MNRAS.484.5453C,2020MNRAS.494.4291C,2022MNRAS.511.2610C} which we illustrate in the figure.

Upon examining the plot, we see that the mean stellar mass~(within the $16$th to $84$th quantiles) lies above the abundance matching relation for a given DM mass. This suggests that our galaxies have similar characteristics to the barred galaxies observed in the Auriga simulations~\citep{2021A&A...650L..16F}. However, when we divide the galaxies into different groups based on their $\rm A_2$ values~(as shown in the right panel), we do not observe substantial differences between strongly barred galaxies (indicated by reddish colours) and axisymmetric galaxies (indicated by bluish colours). Moreover, non-barred galaxies appear slightly more dominated by baryonic matter, but their stellar mass range for a given DM mass is roughly the same. Given that both the TNG50 and Auriga simulations utilize essentially the same subgrid galaxy formation model, incorporating effective stellar wind models~\citep{2013MNRAS.436.3031V, 2014Natur.509..177V,2018MNRAS.473.4077P,2017MNRAS.465.3291W}, it appears that the high baryon fraction is a common feature of the MW and M31 analogues in the TNG50 simulation, but it does not seem to be the governing factor favouring the formation of bars. Interestingly, in Fig.~\ref{fig::st2dm}, the DM-to-baryon fraction of the MW is matched better by galaxies with strong bars in TNG50 but there is an offset from their non-barred counterparts of the same DM mass.

\begin{figure}
\begin{center}
\includegraphics[width=1\hsize]{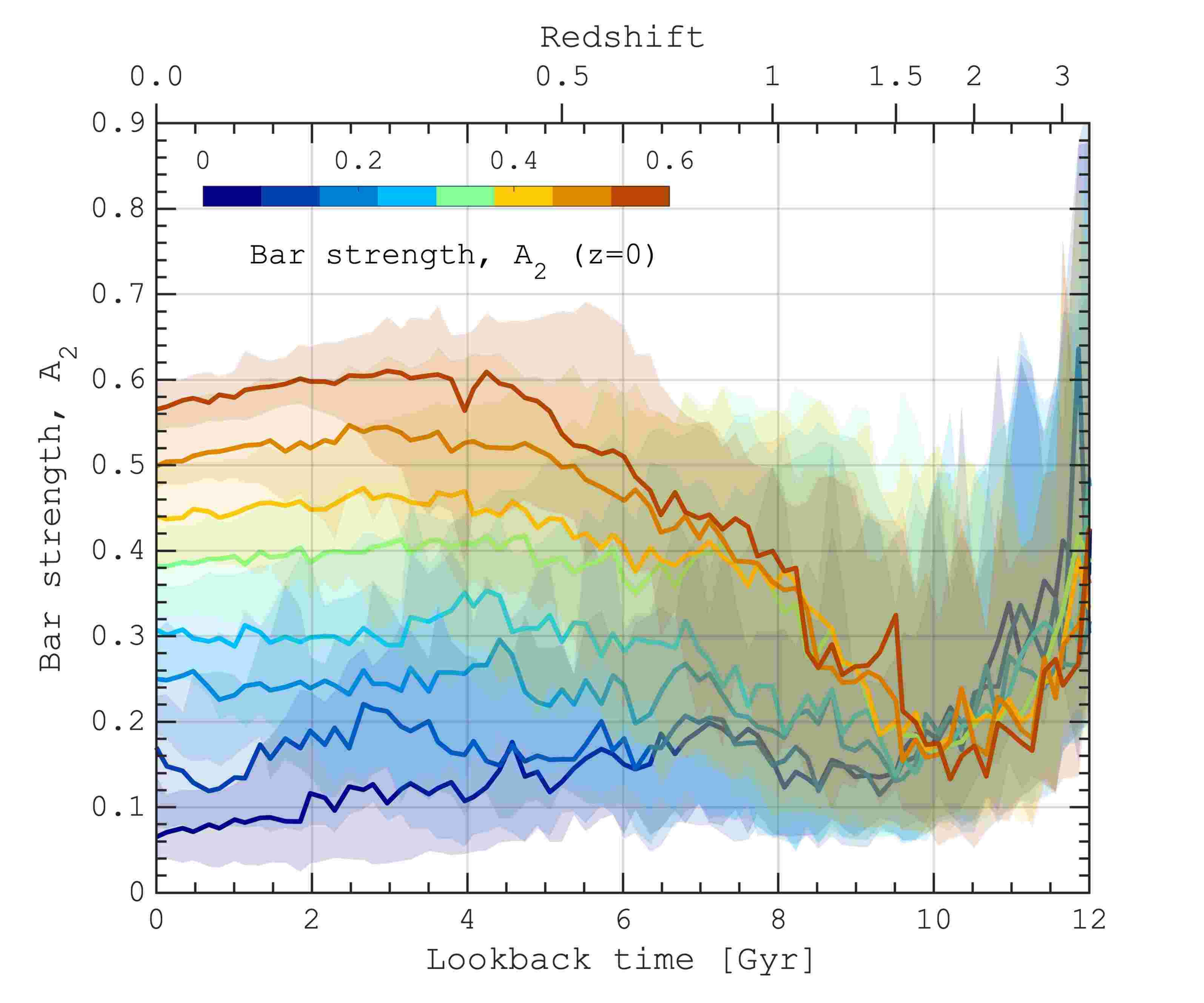}
\caption{Evolution of the bar strength $A_2$ averaged in bins according to their value at $z = 0$~(solid lines), as marked in the colour bar. The filled areas correspond to 16th-84th percentiles. On average, all the bars start to grow steadily at redshift $\approx 2$, where the rate of growth correlates with the maximum strength. We note that there is a number of galaxies with strong bars $A_2>0.3$ at redshift $z>2$. There is a declining trend for the strongest bars, likely due to the formation of pseudo-bulges.}\label{fig::bar_strength_evolution}
\end{center}
\end{figure}

\subsection{Evolution of bar strength with cosmic time}

Figure \ref{fig::bar_strength_evolution} depicts the evolution of the bar strength $\rm A_2$ averaged in bins based on their values at $z=0$. We note that at high redshifts (approximately $z>2.5-3$), the identification of truly barred galaxies in automatic mode may be prone to errors due to the intense accretion of satellites and massive mergers, which can cause strong non-axisymmetric structures in galaxies. Nevertheless, upon visual confirmation, a number of strong bars can be found during that period. We refer to \cite{2022MNRAS.512.5339R}, where the bars fraction evolution in TNG50 simulation is discussed in more detail, showing a reasonable trend with redshift. 

At redshift $z\approx2$, galaxies in all $\rm A_2$ bins (measured at $z=0$) transition to the most axisymmetric morphology with $\rm A_2\approx 0.2$. Subsequently, the bars in all bins begin to grow steadily over time. Interestingly, the averaged strength of bars correlates with the growth rate between redshift $\sim 0.4$ and $\sim 2$. We stress, however, that since we analyse the averaged trends, this should not be considered as a general rule in individual galaxies. Later the bar strength saturates and even diminishes for the strongest bars, likely due to the emergence of pseudo-bulges, which are seen in many TNG50 galaxies at $z=0$~\citep{2023arXiv230212788A}. This phenomenon has been observed in various isolated simulations~\citep{2004ApJ...613L..29M,2006ApJ...637..214M, minchev12a,2013MNRAS.429.1949A}. However, in these tailored models, bars tend to give rise to X-shaped or boxy-peanut bulges relatively soon after their formation~(on a scale of $<1-2$ Gyr). Therefore, the potentially recent emergence of boxy-peanut pseudo-bulges in the MW-type barred analogues is intriguing, particularly since we lack any constraints regarding the timing of its formation in the MW. While it is true that we observe instances of bar destruction and, in certain cases, their subsequent re-formation in individual galaxies, Fig.~\ref{fig::bar_strength_evolution} shows that, on average, bars tend to be old structures. Additionally, non-barred galaxies at $z=0$ rarely exhibit bars at early epochs. 

\begin{figure}
\begin{center}
\includegraphics[width=1\hsize]{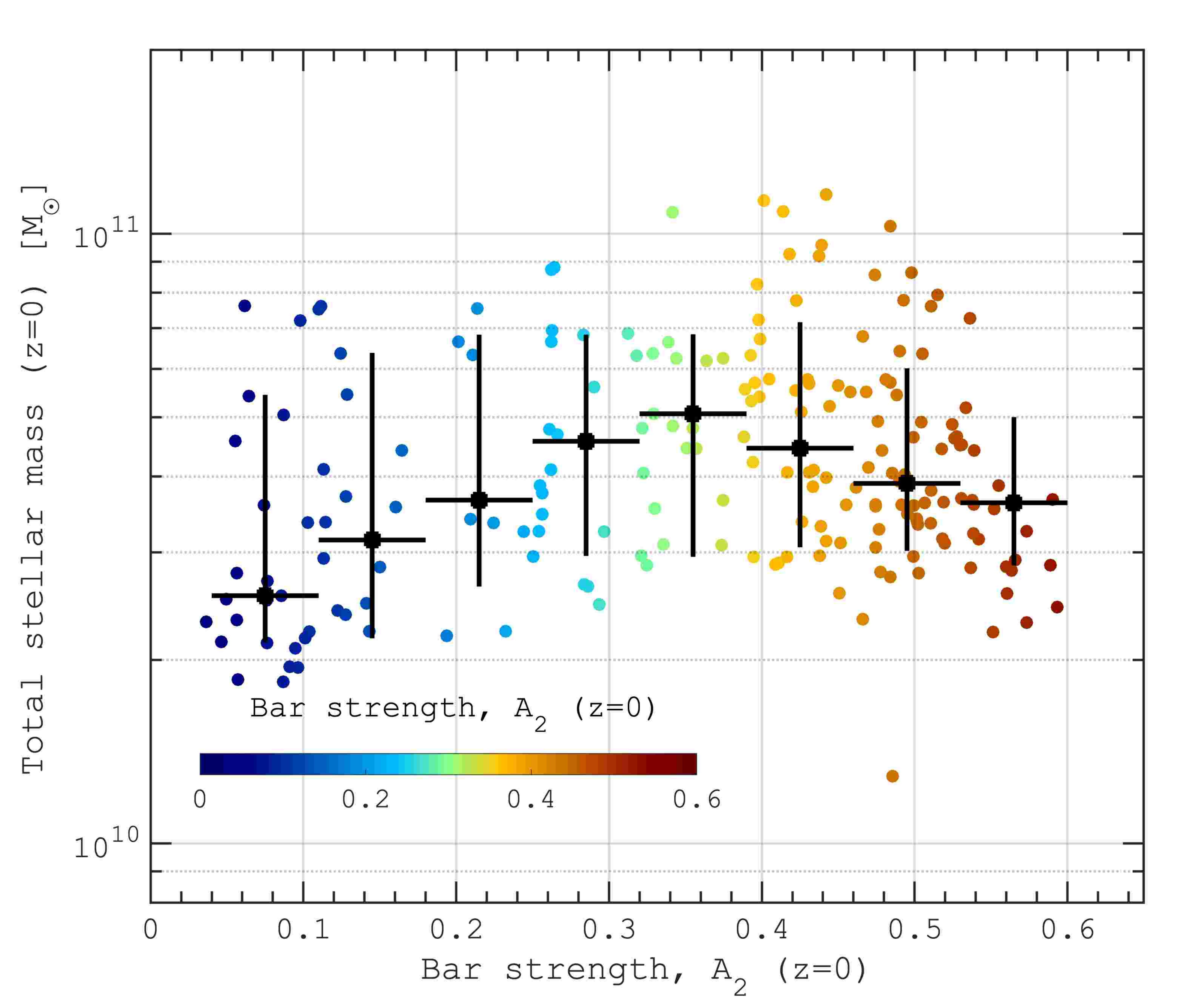}
\caption{Disc mass-bar strength relation at $z=0$. The black symbols correspond to the stellar mass averaged in bins~(shown by horizontal error bars) of according to the $A_2$ and the vertical error bars show 16th to 84th percentiles. }
\label{fig::bar_strength_vs_mass_z=0}
\end{center}
\end{figure}

In Figure \ref{fig::bar_strength_vs_mass_z=0}, we examine the relationship between stellar disc mass and bar strength at $z=0$. Interestingly, we observe a positive correlation in the mean (black dots) up to a bar strength of approximately $A_2 \approx 0.35$, after which a downturn is found. A large scatter in disc mass exists for a given bar strength. The strongest bars are not hosted by the most massive discs, but by those of intermediate and even lower mass. This discrepancy suggests that the relationship between bar strength and disc mass at $z=0$ is not a straightforward linear correlation~\citep[see also][]{2022MNRAS.514.1006I}, which we have already noticed in Fig.~\ref{fig::st2dm}, and other factors may come into play indicating potential disparities in the formation histories of barred and non-barred galaxies, a topic that we delve into further in the subsequent sections.

\begin{figure*}
\begin{center}
\includegraphics[width=0.5\hsize]{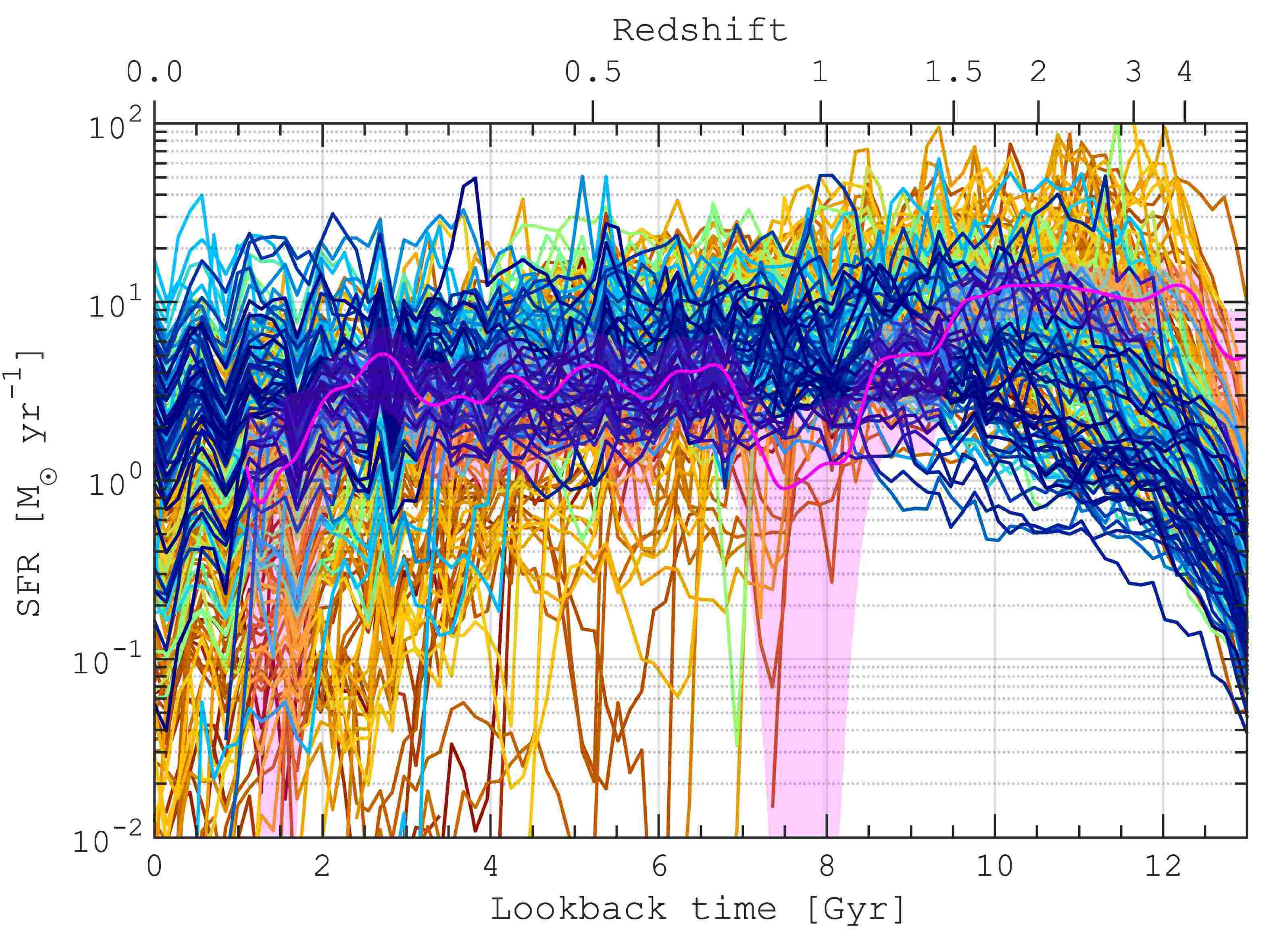}\includegraphics[width=0.5\hsize]{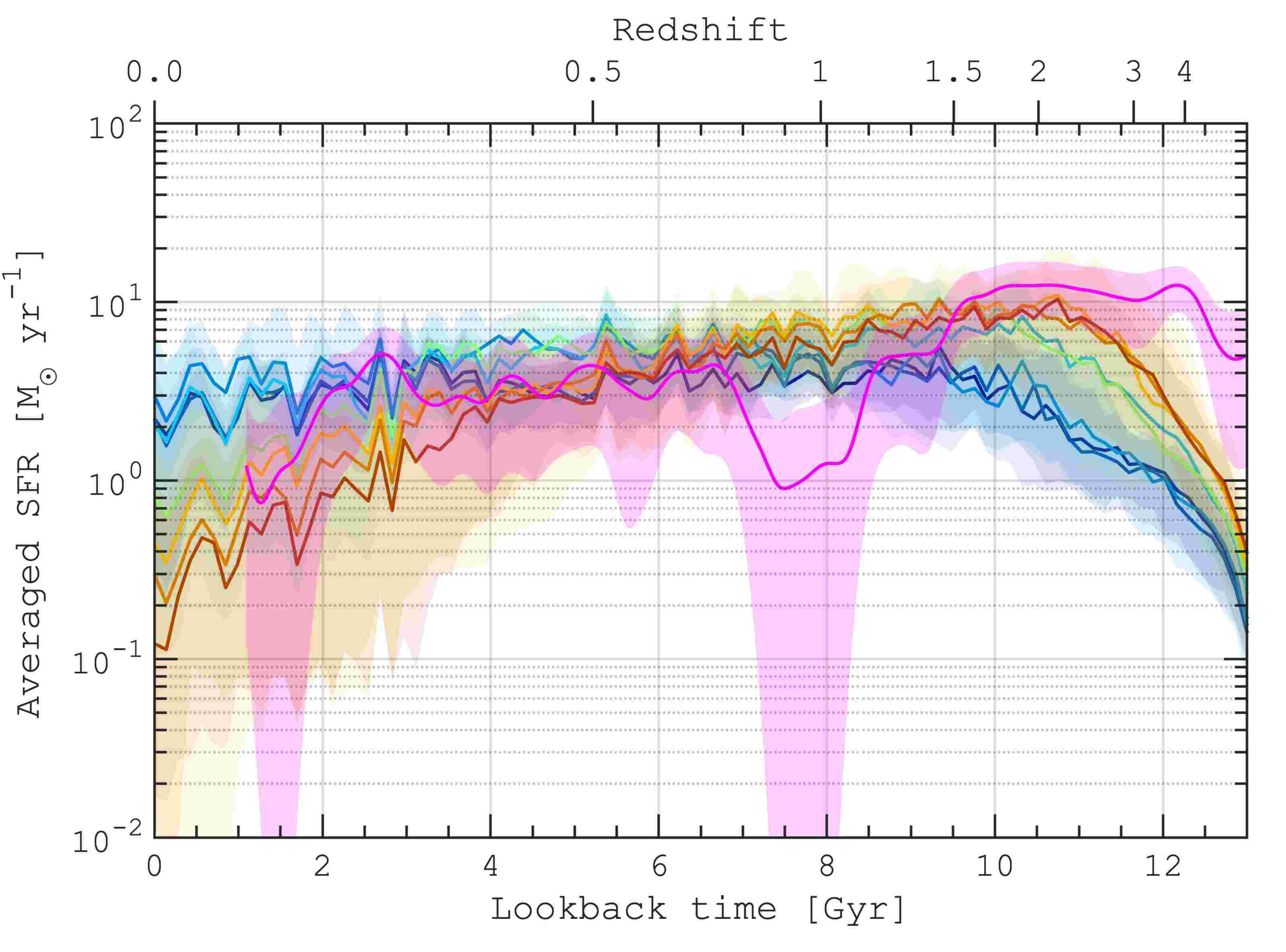}\\
\includegraphics[width=0.5\hsize]{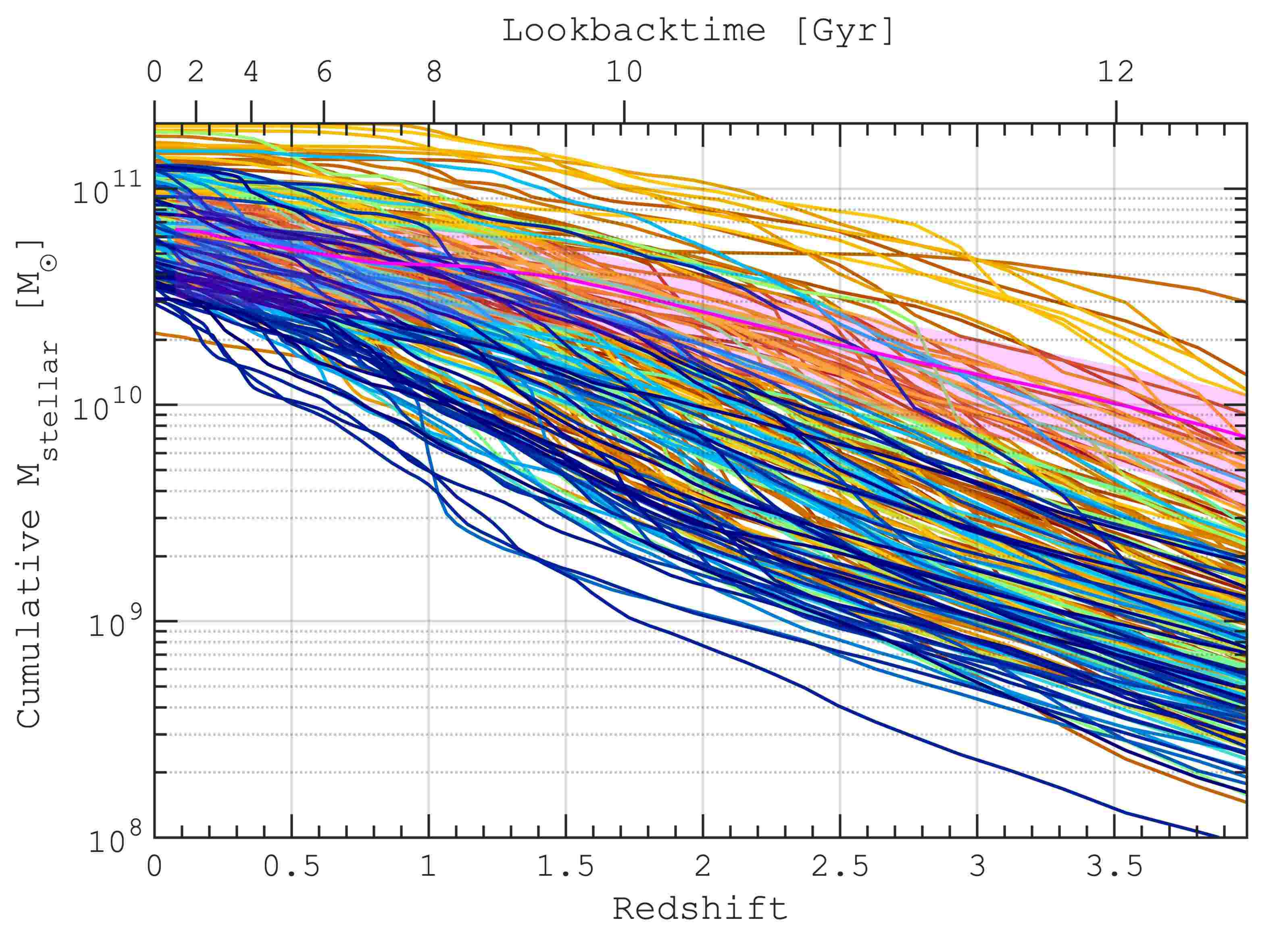}\includegraphics[width=0.5\hsize]{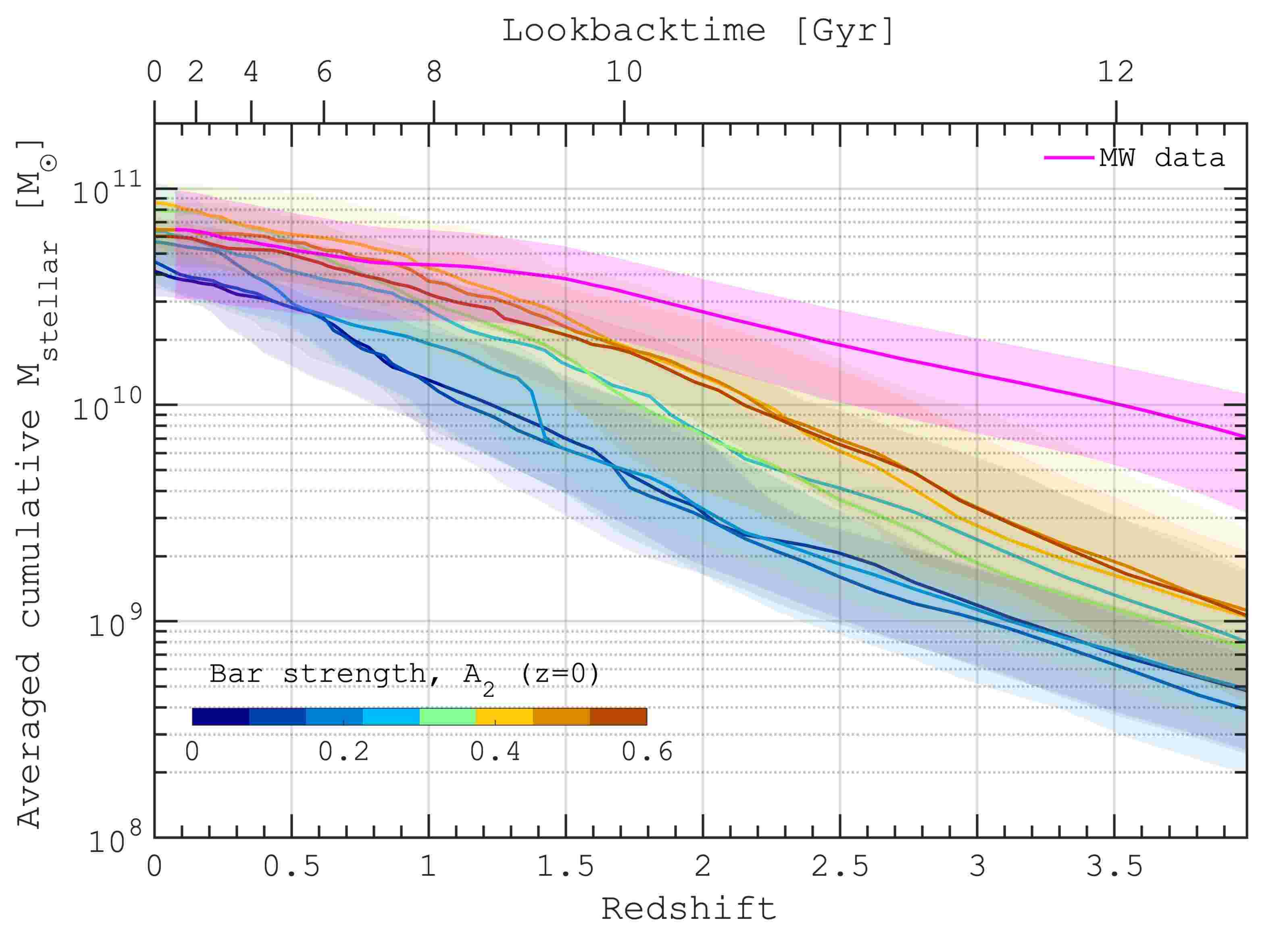}
\caption{Star formation histories of MW and M31 analogues in TNG50 vs the MW. The top panels show the star formation histories of TNG galaxies in our sample colour-coded according to the bar strength~($A_2$) at $z=0$. The bottom panels show a cumulative stellar mass growth as a function of redshift. The left panels display individual star formation histories, while the right ones correspond to averaged values in bins according to the $A_2$ value at $z=0$, as marked by the colour bar. The MW star formation history~(magenta) with error bars are adopted from \protect\cite{2015A&A...578A..87S}. }\label{fig::sfhs}
\end{center}
\end{figure*}

\begin{figure*}
\begin{center}
\includegraphics[width=1\hsize]{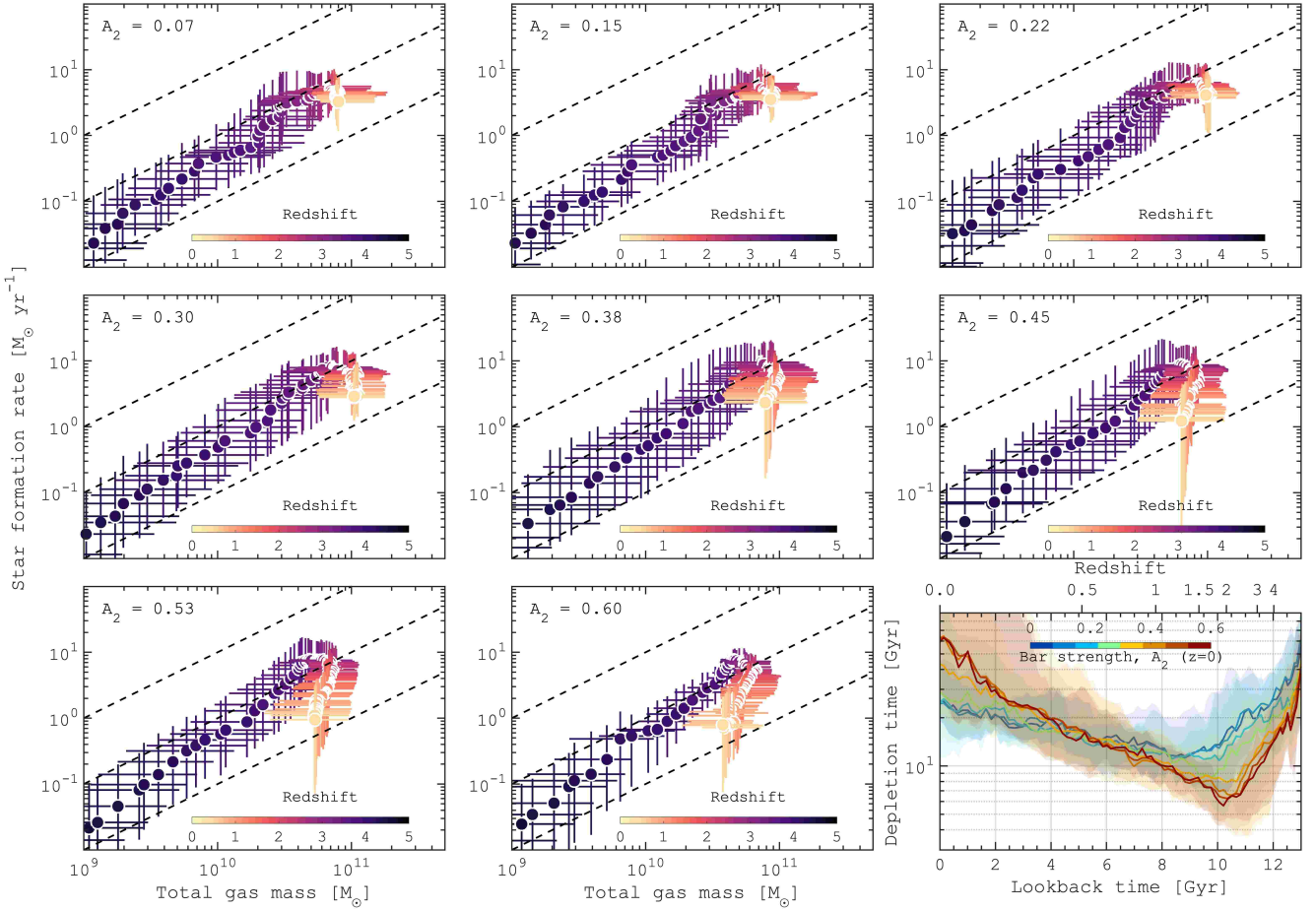}\caption{Relation between total gas amount and total star formation rate as a function of redshift for galaxies in different bins of the bar strength~($\rm A_2$) measured at $z=0$. The averaged values are shown with circles, and the error bars depict the 16th-84th quantiles for a given time. The bottom right panel shows the evolution of the averaged depletion time~(ratio between the gas mass and SFR, in Gyr) as a function of time for galaxies in different bins of $A_2$ at $z=0$. The shaded areas show the 16th-84th quantiles of the depletion time. We stress that since we use the total gas amount inside the haloes, the depletion times are longer than it is expected if only gaseous discs or molecular gas is considered. }
\label{fig::sfe}
\end{center}
\end{figure*}

\subsection{Stellar mass and star formation evolution}

We next access the mass assembly history of TNG50 galaxies. In Fig.~\ref{fig::sfhs} we show the star formation histories and cumulative stellar mass growth for individual galaxies and averaged in bins according to the bar strength. A striking feature is a clear trend between the mass growth rate and the bar strength. This behaviour for barred and non-barred galaxies has already been noticed in \cite{2022MNRAS.512.5339R}, in agreement with MANGA galaxies~\citep{2020MNRAS.495.4158F}. This suggests that barred galaxies formed faster compared to non-barred ones, which accumulate their stellar mass on a longer time scale. Indeed, at early epochs~($z>1-1.5$) barred galaxies experience an intense starburst phase, while non-barred ones acquire their stellar mass in a more quiescent manner. 

In Fig.~\ref{fig::sfhs}, the star formation rate (SFR) in barred galaxies is by an order of magnitude lower compared to non-barred galaxies at low redshifts, and there is a rather smooth transition from one to the other. This picture seems to be in agreement with observational data of external barred galaxies, which show a deficit of gas and lower SFR at low redshifts~\citep{2017A&A...607A.128G,2020MNRAS.495.4158F}.  Figure \ref{fig::sfhs} clearly demonstrates that barred galaxies genuinely grow faster at high redshift. However, at lower redshifts, the bars themselves can affect the star formation histories (SFHs). For instance, the SFR of barred galaxies starts to decline at around the same time when bars start to grow, at $z=1.5-2$. It is well known that bars redistribute gas, thus affecting star formation~\citep{1983IAUS..100..215P,1992MNRAS.259..345A,2015MNRAS.454.3299R, 2018A&A...609A..60K} which brings us to the question of the amount of gas available for star formation and their evolution depending on galaxy morphology.

In Figure \ref{fig::sfe}, we present the redshift evolution of a relationship between the total gas mass inside the galactic halo and the total star formation rate (averaged over the previous $100$ Myr) for all galaxies, grouped according to their final bar strength~(different panels correspond to different bar strength, as marked in the top left of each panel). This figure allows us not only to trace the evolution of gas mass and SFR but also to assess the star formation efficiency, highlighted by diagonal lines representing constant depletion times (in Gyr). The evolution of the averaged depletion time is shown in the bottom right panel of the figure. We stress that the depletion times are much larger than it is typically found because we calculate them as a function of the total gas amount, not considering only the gas inside the disc or its molecular phase density. 

In Fig.~\ref{fig::sfe} we see almost a monotonic~(power-law) relation between the SFR and the amount of gas for non-barred galaxies~(top two panels). This suggests the gas mass directly regulates the rate of star formation. This is also evident from a small variation of the depletion time as a function of lookback time~(bluish lines in the bottom right panel). However, once the bar strength increases a pivot point emerges in the relation. Around this point, the total amount of gas decreases inside the haloes, which can be caused either by its fast consumption by star formation and/or by the AGN feedback resulting in a quenching of star formation, but this is seen more prominently in the barred galaxies. In both cases, the amount of gas decreases either by its consumption or by its removal from the galaxy. It was demonstrated that the impact of the  black hole kinetic wind feedback is responsible for the inside-out quenching of discs in the TNG50~\citep{2021MNRAS.508..219N}; however, the correlation between bars and the strength of quenching is quite remarkable. While the exact sequence of these mechanisms extends beyond the scope of our study, it seems that a bar-triggered cycle likely plays a significant role in this process, contributing to the observed diversity of barred and non-barred galaxies at low redshifts. 

What is even more intriguing in the context of our study is that barred galaxies show a higher star formation efficiency at $z>3-4$, well before the epoch of the formation of the bars ~($z\approx 2$, see Fig.~\ref{fig::bar_strength_evolution}). Therefore, next, we need to assess the physical conditions at the very early epochs, favouring a higher star formation efficiency in galaxies where the bars will develop later.

\subsection{Tracing discs formation}\label{sec::tracing_disks}

A crucial finding of our study is that galaxies which  develop bars exhibited high star formation efficiency during the early stages of evolution~($z\approx 2-4$) with star formation bursts of up to $50-100$~\Msunyr. This is not the case for non-barred galaxies, which show rather quiescent SFHs at high redshift. Therefore, one needs to clarify what physical conditions support a high SFR along with a high star formation efficiency.

A recent study by \cite{2022MNRAS.516.2272S} highlights the role of a galactic disc in enabling highly efficient star formation in the VINTERGATAN cosmological simulation~\citep{2021MNRAS.503.5826A,2021MNRAS.503.5846R}. This work suggests that a coherent motion of gas  within a rotationally-supported~(i.e. discy) galaxy is necessary to sustain a high star formation rate on a Gyr-long time scale~(bursty phase), as it allows for gas concentration within a confined volume and, thus, to form stars efficiently. In contrast, in a more chaotic and dispersion-dominated system, it becomes difficult to envision a prolonged bursty star formation phase, as feedback processes could scatter the material across a larger volume, effectively suppressing further star formation. A slightly different perspective is given in another recent work by \cite{2023MNRAS.tmp.1847H}, who, using a large suite of simulations of dwarf galaxies, demonstrated that disc formation is possible in both bursty and smooth star formation regimes of star formation. Although a number of simulations suggest the link between the transition between star formation activity and galaxy structure~\citep{2017MNRAS.467.2430M,2019MNRAS.484.3476C,2021MNRAS.505..889Y,2021MNRAS.508..352R,2023MNRAS.519.2598G}, unravelling the intricate relationship between bursty star formation and disc formation process poses a challenge. Indeed, it is difficult to determine whether bursty star formation precedes the establishment of a galactic disc or if galaxies that settle into a disc configuration can sustain a high star formation rate over an extended period of time. 

Within our sample, which comprises both bursty galaxies (predominantly barred) and galaxies with a relatively quiet star formation history (predominantly non-barred), we can investigate the formation history of their discs. To quantify the level of discyness, we employ the widely used circularity parameter $\lambda$~\citep{2003ApJ...597...21A}. Instead of using a sharp boundary of $60-70\%$ for the circularity-based disc stars selection, which might not work properly at high redshift, during the bursty phase of star formation, we assume that the circularity distribution of a spheroidal component~(either bulge and/or inner halo) is symmetric around zero value. In other words, we define the spheroidal  component as twice the fraction of stars with $\lambda<0$, and the rest is assumed to be the disc component~\citep{2003ApJ...597...21A, 2014MNRAS.437.1750M}. 

\begin{figure}
\begin{center}
\includegraphics[width=1.0\hsize]{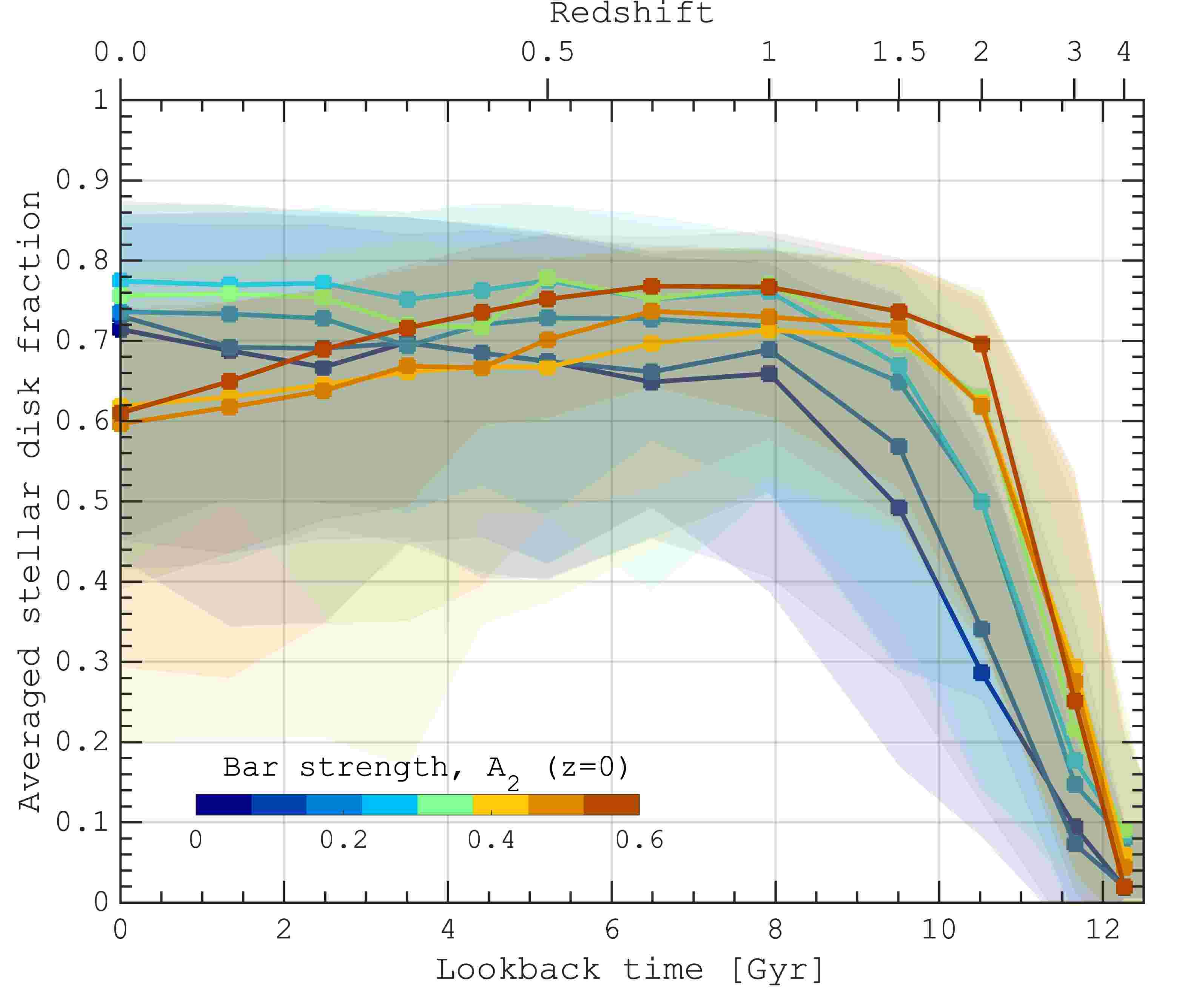}
\caption{Formation of disc components in TNG50 galaxies. The evolution of the disc component fraction for different bins of the bar strength at $z=0$ is shown. The disc fraction is calculated using a circularity parameter, see Sec.~\ref{sec::tracing_disks} for more details.}\label{fig::disks}
\end{center}
\end{figure}

In Fig.~\ref{fig::disks} we show the averaged evolution of the circularity-defined stellar disc fraction for galaxies in different bins of the bar strength at $z=0$. Diverse formation paths, already observed in the star formation history, are also seen in the discs growth. In particular, discs in barred galaxies tend to form faster, at $z\approx 2$, discs represent about $70\%$ of the stellar mass, while non-barred galaxies reach such values at $z\approx 1$. Since the barred galaxies are more massive at early times, a larger fraction of discs means that the discs of barred galaxies are already quite massive $1-2 \times10^{10}\Msun$ at $z>1$. This striking diversity in the evolution of galaxies as a function of the bar strength suggests that the early emergence of discs promotes the formation of bars. This result is trivial because, although the exact conditions are uncertain~\citep{2023MNRAS.518.1002R}, the bar instability can only develop in disc-dominated systems~\citep{1981A&A....96..164C,1982MNRAS.199.1069E,1993RPPh...56..173S}. However, this brings us back to a question already raised in the introduction: why do not all disc galaxies host bars?

As we have seen, TNG50 barred galaxies need only about $1-2$~Gyr to form prominent bars~(see Fig.~\ref{fig::bar_strength_evolution}). In this case, there exists ample time for~(non-barred) galaxies that are gradually growing to eventually undergo a bar instability, given that they transition into a disc-dominated state at $z=0.5-1$~(see Fig.~\ref{fig::disks}). Apparently, the answer comes from the selection procedure of the MW/M31 analogues. Since one of the criteria is the total stellar mass of galaxies at $z=0$, if some (non-barred) galaxies did not acquire enough stellar mass at early epochs, in order to arrive at the desired MW~(or M31)-like stellar mass at $z=0$, they need to sustain a high SF rate at lower redshifts, which implies a substantial amount of either in-situ gas or its recent infall. However, as we know, a relatively high gas fraction suppresses the bar instability~\citep[see, e.g.][]{2005MNRAS.364L..18B,2013MNRAS.429.1949A}. Therefore, the bars can form in early discs that already lack a substantial amount of gas efficiently converted into stars. 

We see that discs in some TNG50 galaxies formed early, but in some cases, for some reason, the process of disc assembly is delayed. For a long time, the governing mechanism of disc formation has remained enigmatic due to the complexity and multiplicity of different processes poorly constrained by observational data at high redshifts. Nowadays, it is believed that the formation of galactic discs is attributed to the conservation of angular momentum during the collapse of gas within extended dark matter haloes, where the angular momentum can be acquired through mergers as well. In this scenario, a filamentary accretion of gas with high angular momentum can set up the discs~\citep{2012MNRAS.423.1544S,2012MNRAS.422.1732D,2015MNRAS.449.2087D, 2013ApJ...769...74S, 2020MNRAS.497.4346K,2022MNRAS.510.3266K}.  This theoretical picture is in agreement with recent observations that push the time scale of discs formation to $z>4$~\citep{2020Natur.581..269N, 2020Natur.584..201R, 2021A&A...647A.194F, 2021Sci...371..713L}. 

\begin{figure*}
\begin{center}
\includegraphics[width=1.0\hsize]{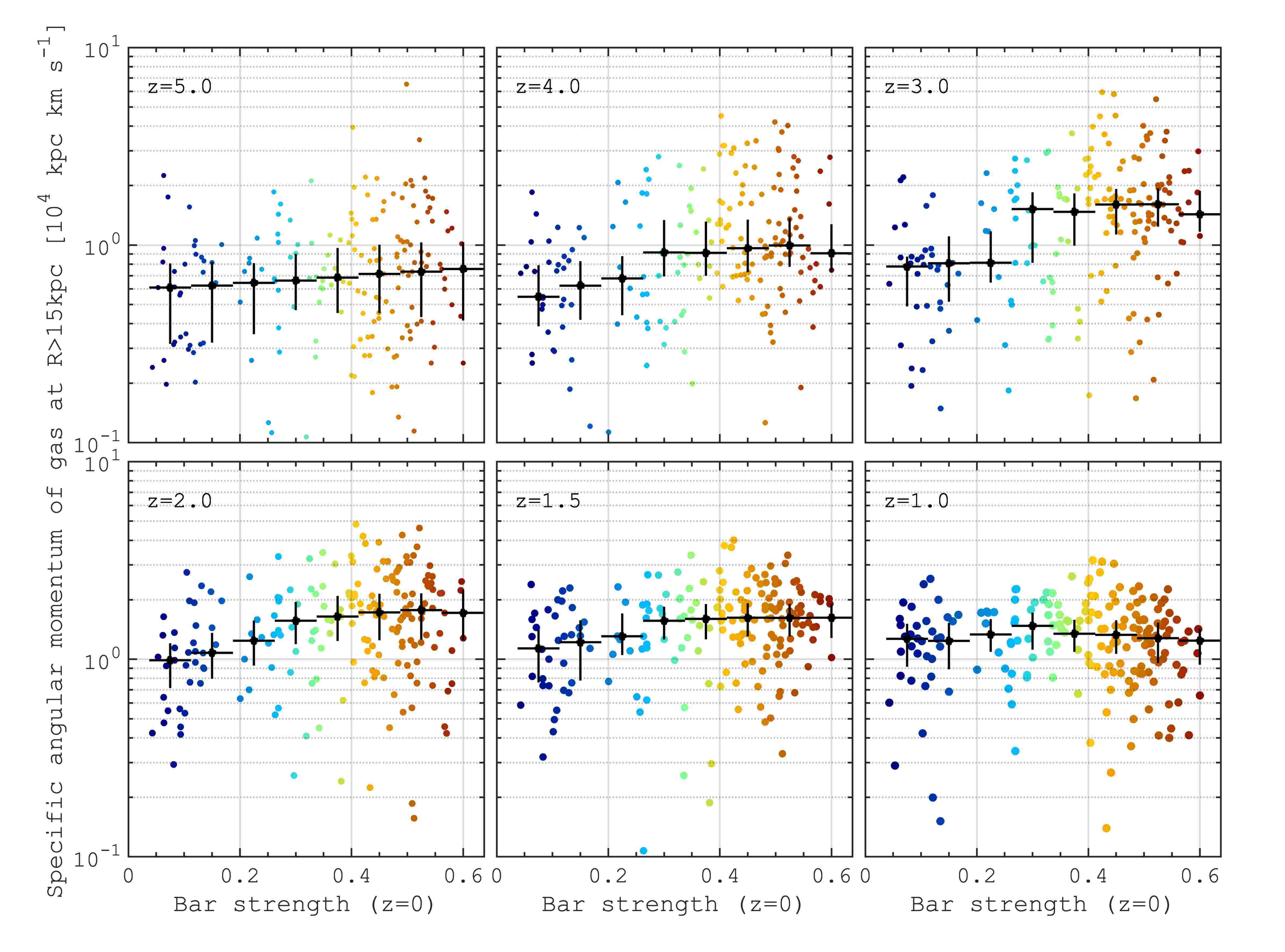}
\caption{Evolution of the specific angular momentum of gas~($j_{gas}$) surrounding TNG50 galaxies as a function of the bar strength at $z=0$. The averaged values of $j_{gas}$, calculated outside a sphere of $15$~kpc, are shown by the black symbols where the vertical bars correspond to $16$th-$84$th quantiles. 
}
\label{fig::AM}
\end{center}
\end{figure*}

Following the above-mentioned observational and theoretical results, we test the diversity of the conditions for the formation of discs of barred and non-barred galaxies. 
In Fig.~\ref{fig::AM} we present the relation between the specific angular momentum of gas at $z=5, 4, 3, 2$ and $1$ and the bar strength at $z=0$. In order to avoid potential contamination from proto-discs, we calculate the angular momentum of gas outside the sphere of $15$~kpc. Already from $z=5$, gas around proto-barred galaxies has a higher angular momentum. The spread in the averaged angular momentum for strongly barred and non-barred galaxies is the largest at $z=3$, but this difference almost washed out at $z=1$. We note that between $z=1-3$ the averaged angular momentum of gas in proto-barred galaxies remains constant, while for non-barred ones, it slowly increases and reaches similar values. This figure illustrates why some discs grow faster and form bars, while others experience slower disc formation. For low angular momentum systems, the orientation of the gas infall is not coherent, resulting in dispersion-dominated systems~\citep{2012ApJS..203...17R,2015ApJ...812...29T,2015ApJ...804L..40G,2016MNRAS.463..170C,2018ApJ...868..133F} where, as we already discussed, it is hard to maintain a high rate of star formation on a long time scale. At the same time, a coherent flow, preferably from filaments at high redshift, regulates the rate of rotation of (proto)-galaxies~\citep{2014MNRAS.444.1453D, 2013MNRAS.428.2489L, 2013ApJ...775L..42T, 2016MNRAS.457..695P, 2018ApJ...866..138W, 2021NatAs...5..839W}. Therefore, our analysis suggests that the correlations between the early discs settling and subsequent bars formation in the TNG50 galaxies can be understood as a consequence of the dynamics of large-scale cosmic flows.

\begin{figure*}
\begin{center}
\includegraphics[width=0.5\hsize]{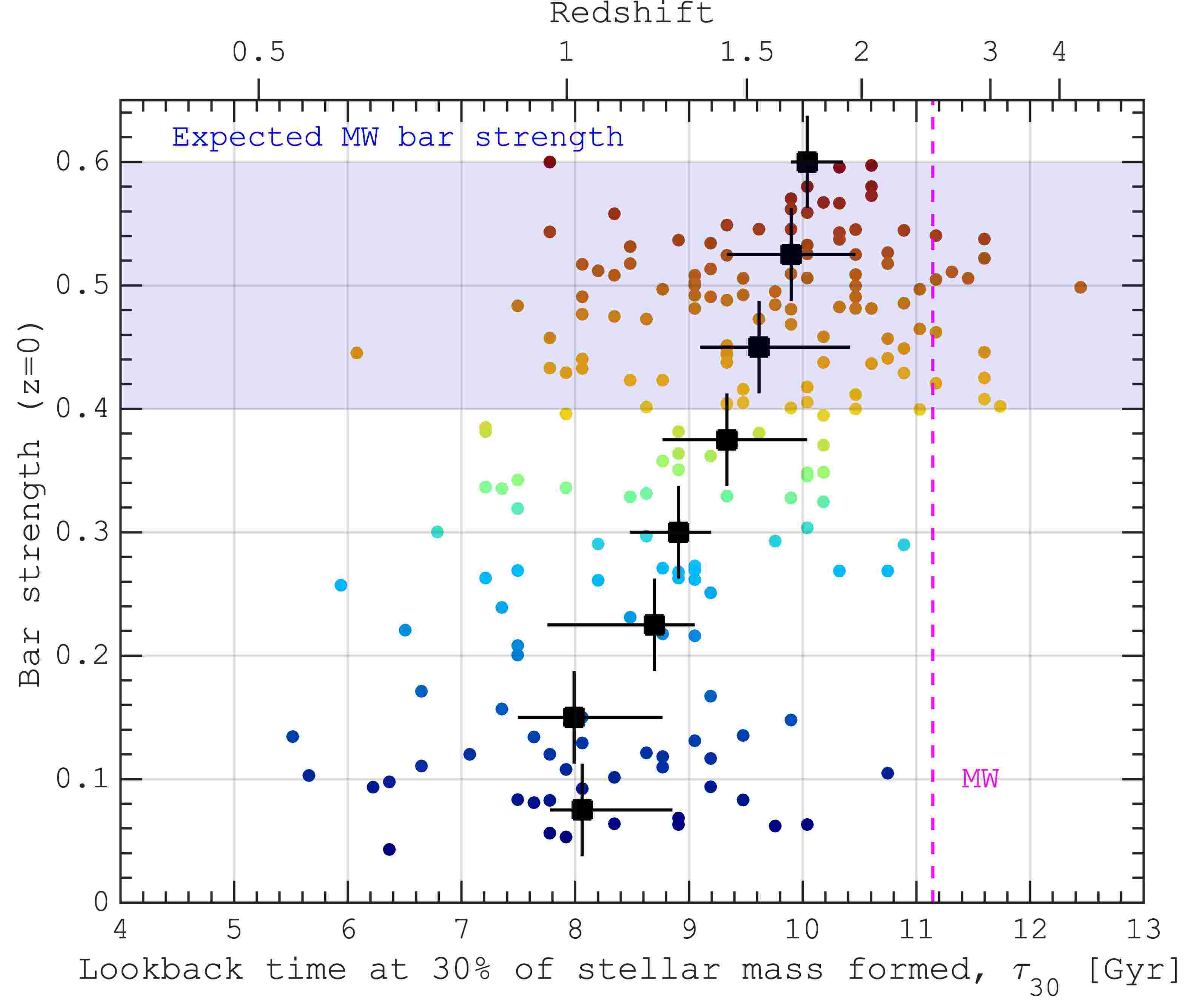}\includegraphics[width=0.5\hsize]{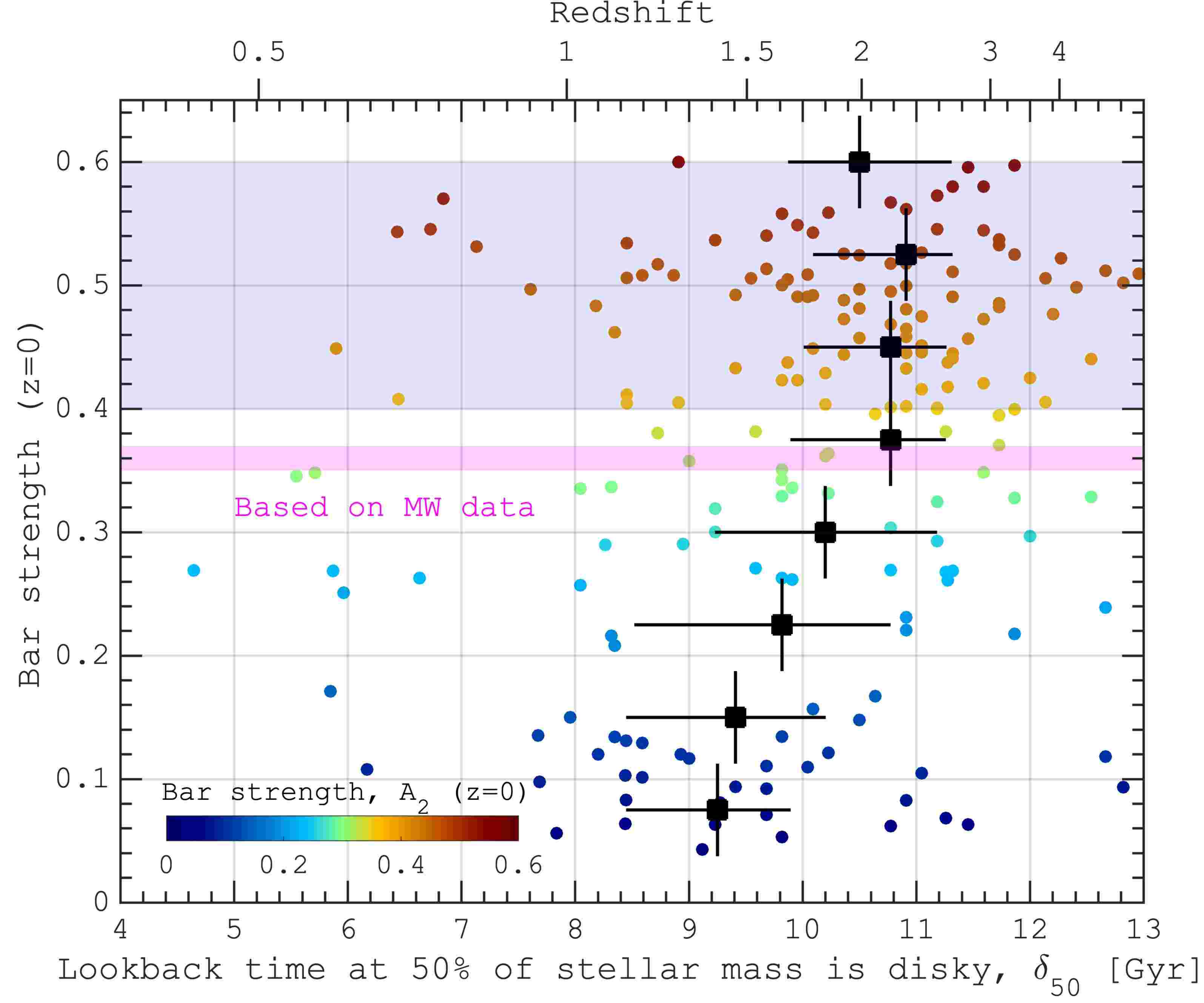}
\caption{MW bar and disc formation inferred from the TNG50 analogues. {\it Left:} relation between the bar strength, $A_2$ at $z=0$, and the lookback time when $30\%$ of the total stellar mass formed. {\it Right:} relation between the bar strength, $A_2$ at $z=0$, and the lookback time when $50\%$ of the stellar mass has a discy kinematics, according to the circularity parameter distribution. In both panels, the blue filled area corresponds to the expected MW bar strength from the comparison of the MW stellar mass growth and the TNG50 analogues. In the right panel, the magenta line corresponds to the directly estimated MW bar strength. The values averaged in different $A_2$ bins are shown by the black symbols where the horizontal bars correspond to $18$th-$84$th quantiles. }\label{fig::mw_params}
\end{center}
\end{figure*}

\section{Implications to the Milky Way}\label{sec::MW-bar} 

As we showed above, a number of galaxy properties in the past correlate with the present-day strength of the bar. Hence, we can use the strength of the MW bar to put constraints on the evolution of the Galaxy. The MW star formation history from \cite{2015A&A...578A..87S} already suggests that the Galaxy must have a bar, as it behaves like a typical barred galaxy from the TNG50 sample~(see Figs.~\ref{fig::st2dm} and \ref{fig::sfhs}), with a rapid early mass growth when, as suggested in the literature~\citep{2013A&A...560A.109H,2014ApJ...789L..30L,2015A&A...578A..87S,2018A&A...618A..78H}, a massive high-$\alpha$~(or inner thick disc) component of the MW should have been formed~\citep[see][ for simulations results]{2018MNRAS.474.3629G,2021MNRAS.501.5176K,2020MNRAS.491.5435B,2021MNRAS.505..889Y,2021MNRAS.503.5868R}. 

Since the bar formation is the result of stellar disc instability, the age of the MW bar sets a lower limit for the MW disc formation epoch. Similarly to the TNG galaxies, where the bar formation epoch correlates with the end of the bursty high-efficiency SF~(see Fig.~\ref{fig::bar_strength_evolution}), we expect that the MW bar has begun to grow at a redshift of $1.5-3$. One way to test this is the age of stars in the nuclear stellar disc of the MW, which, as theory suggests, started to form at the time of the bar formation \citep{2014MNRAS.445.3352C, 2020MNRAS.492.4500B}.  Models of chemical abundance trends suggest that some of these stars should be as old as $8-12$~Gyr \citep{2020ApJ...894...26T}, which is also shown in several recent studies~\citep{2020NatAs...4..377N, 2023A&A...671L..10N, 2023A&A...672L...8S}, allowing even to put tight constrains on the MW bar formation time $8\pm 1$~Gyr ago~\citep{2024MNRAS.tmp..735S} manifested by a significant burst of star formation in the nuclear disc.

Another promising concept to put a constraint on the time of the MW bar growth relies on the migration of high-metallicity low-eccentricity stars from the galactic center~\citep{2020A&A...638A.144K}. In this case, stars, trapped by the resonances, transferred towards the solar radius due to the slowdown of the bar, which, as $N$-body models suggest, happens rapidly after the beginning of the bar formation, dating the epoch of the intense bar growth. Current age information about the super-solar metallicity populations and their spatial distribution in the inner Galaxy implies the MW bar formed around $7-10$~Gyr ago~\citep{2023A&A...669A..96D, 2024arXiv240308963H}, in agreement with the nuclear stellar disc data~(but see alternative claims in \cite{2024A&A...681L...8N}).


We therefore conclude that different methods arrive to similar conclusions suggesting that the MW bar started to form around redshift $0.8-1.8$, as in the TNG galaxies, which would not contradict observational data about external galaxies~\citep{2015A&A...584A..90G, 2023A&A...671A...8D}. We note that in the case of bars with a moderate strength, which grow relatively slowly~(see Fig.~\ref{fig::bar_strength_evolution}), the formation of nuclear discs can be delayed to the time when the bar strength reaches relatively high values~\citep{2014MNRAS.445.3352C}.
 
In the following, using the available data about the MW, we attempt to infer some parameters which are in favour of its bar formation. In Fig.~\ref{fig::mw_params} we show a correlation between the lookback time when $30\%$ of stellar mass formed for all TNG galaxies as a function of the bar strength at $z=0$. As expected from the SFHs~(see Fig.~\ref{fig::sfhs}), there is a nice correlation where the horizontal scatter is likely caused by diverse mergers and interactions history, feedback and gas inflow cycles. The star formation history from \cite{2015A&A...578A..87S} suggests that the MW acquired $30\%$ of its mass around $11.2$ Gyr ago, which is marked by the vertical magenta line. We note that the choice of $30\%$ mass threshold for this figure is somewhat arbitrary; however, from Fig.~\ref{fig::sfhs}~(bottom left), it is evident that the MW follows the mass assembly of galaxies with the strongest bars. The MW-based data vertical line crosses the tail of the TNG-based relation for galaxies with the bar strength of $0.4-0.6$. Therefore, the mass assembly history not only suggests that the MW should host the bar but also gives an idea regarding its strength. 

Unfortunately, there are no direct measurements of the MW bar' strength in units of $\rm A_2$. For this reason, in Fig.~\ref{fig::bar_strength}, we presented a correlation between $\rm A_2$ and major-to-minor axes density ratio. The latter one also cannot be measured directly even with the most recent \Gaia~(DR3) data; however, there are several models based on various observations of the inner MW, which provide some estimates of the density distribution axes ratio in the bar region. \cite{1997ApJ...477..163S} and \cite{2013MNRAS.434..595C} found the axes ratio of $0.43$~($B_{s} \approx 0.57$ and $A_2\approx0.35$); more recent measurements  by \cite{2013MNRAS.435.1874W}, by taking into account a triaxial model of the bulge, reveal roughly the same in-plane axes ratio of $0.44$~($B_{s}\approx 0.56$ and $A_2 \approx 0.37$). A problem with such estimates is the presence of massive~~\cite[$\approx 1.85\times 10^{10}\Msun$,][]{2013MNRAS.435.1874W} pseudo-bulge, which, while being elongated along the bar, broadens the total density distribution along the minor axis of the bar. Therefore, our estimates of $A_2\approx 0.35-0.37$ from the above-mentioned studies can be used as lower limits for the genuine MW bar, as we do not correct for the effects of the triaxial pseudo-bulge on the density distribution. 

Another promising tool to obtain the strength of the MW is measuring the kinematics of stars in the inner MW~\citep{Soumavo}, where a well-known butterfly-like radial velocity pattern is observed~\citep{2019MNRAS.490.4740B,2021A&A...656A.156Q,2023A&A...674A..37G}. Unfortunately, this region of the MW is still poorly covered by the \Gaia RVS sample, especially in the Galactic midplane due to high extinction. Also, the velocity pattern is obscured by relatively large distance errors~\citep{2023A&A...674A..37G,2024MNRAS.528.3576V} and affected by the bar-spirals connection, complicating estimations of the bar strength~\citep{2020MNRAS.497..933H,2024MNRAS.528.3576V}.

We can summarize that the TNG-based stellar mass growth suggests the bar strength of $A_2\approx 0.4-0.6$ while some indirect but MW-based measurements favour slightly lower values of $A_2\approx 0.35-0.37$. Fortunately, in Fig.~\ref{fig::disks} we notice that galaxies with the strongest bars~($A_2>0.3$, reddish colours) experience very similar stellar disc growth rates. To quantify this, in Fig.\ref{fig::mw_params}~(right), we show the relation between the bar strength and the lookback time when the disc component represents $50\%$ of stellar mass, which is a transition from a spheroid- to a disc-dominated regime. The figure suggests that the TNG50 galaxies, to form a strong~($A_2>0.35$) bar, should be disc-dominated at $z \approx 2$. In the context of the MW, our estimations of the MW bar strength suggest that the MW should have become disc-dominated around the same time, $10.5-11$~Gyr ago, with the mass of the disc-component of $\approx (5-15)\times10^{9}$~\Msun\ at that time. These estimates are in agreement with studies by \cite{2012ApJ...758..106K} and \cite{2017ApJ...843...46S} demonstrating that a star-forming MW-like disc galaxy was likely supported by dispersion at $z>2$, at a time when it was also much less massive.

\section{Discussion}\label{sec::discus}

From the analysis of the MW and M31 analogues from the TNG50 simulations, we concluded that present-day bars trace the evolution of their host galaxies, and in particular, not only their mass assembly but also the disc formation. In this context, the MW, being a typical barred galaxy, is expected to be a disc-dominated system at $z\approx 2$. Thanks to the advance in the massive determination of stellar ages in large MW surveys, we are now confident that this is the epoch when the so-called high-$\alpha$ sequence formed in a relatively short but intense episode of star formation~\citep{1992ApJ...391..651B,1999ApJ...515..226C,2013A&A...560A.109H,2014ApJ...796...38N,2014A&A...562A..71B,2015A&A...579A...5H, lu22, ratcliffe23}. The bursty star formation regime at the time of its formation implies that the disc should have been formed hot, as suggested in many recent works~\citep{brook04,  2013ApJ...773...43B, minchev13, 2014MNRAS.443.2452M, 2021MNRAS.505..889Y, 2022MNRAS.512.3806V}.

The evidence of the MW disc presence at such early times also emerges from the analysis of the impact of the last significant merger~(Gaia-Sausage-Encelladus, GSE, \citealt{2018MNRAS.478..611B,2018ApJ...863..113H,2018Natur.563...85H}), which is seen as a driver of the early MW disc heating. Assuming that the GSE merger happened $8-11$ Gyr ago, there should be a disc already in place to be heated and form the population of Plume/Splash stars~\citep{2019A&A...632A...4D,2020MNRAS.494.3880B}. This also suggests that the disc of the MW should be massive enough in order to survive the massive GSE merger at the time of this event~\citep{2022MNRAS.516.5404S}. At the same time, this should allow for some of the in situ born MW stars to maintain a disc-like kinematics, which is observed for a surprisingly high number of very metal-poor stars~\citep{2020A&A...636A.115D,2020MNRAS.497L...7S}.

Another evidence of the early presence of the MW disc is the so-called ``spin-up'' stars~\citep{2022MNRAS.514..689B}, which have the lowest metallicity~(and thus oldest) in situ stars and steeply rising rotational velocity as a function of metallicity~\citep[see also][]{2022arXiv220402989C,2022ApJ...941...45R}. Such behaviour was found in several simulations~\citep{2023A&A...677A..91K}, including the TNG50~\citep{2023arXiv230609398S,2023arXiv230613125S}. In particular, \cite{2023arXiv230609398S} showed that the MW-like TNG50 galaxies show a sharp spin-up feature with a significant variation of the metallicity at which this feature occurs, $-1.2 \leq \FeH \leq -0.5$ which is lower compared to the FIRE and Auriga simulations~\citep{2022MNRAS.514..689B} but in a better agreement with the MW. This suggests that discs in the selected subsets of simulations may form later, with a lower probability~(but not fully excluding) of hosting MW-like bars.

\section{Summary}\label{sec::concl}

In this work, we used MW and M31 analogues from TNG50 simulations to analyse their evolution as a function of the bar strength at $z=0$. It appeared that the emergence of bars is sensitive to the initial conditions of galaxy formation, in which a small change can result in observed morphological differences of TNG50 galaxies. There is no sharp distinction between the evolution of barred and non-barred galaxies, while there is a smooth variation in their past. Below, we highlight our main results.

\begin{itemize}

\item[$\bullet$] We found a correlation between the bar strength at redshift $z=0$ and the assembly of stellar disc mass. Non-barred galaxies tend to have a relatively smooth history of star formation, resulting in moderate star formation rates and a significant amount of gas remaining at $z=0$. In contrast, barred galaxies exhibit a rapid accumulation of stellar mass prior to approximately $z\approx 1$, leading to minimal or no ongoing star formation and a lower gas fraction.

\item[$\bullet$] The formation of discs in barred galaxies starts at an earlier stage, around redshift $z\approx 2-3$; since then, they stabilize as systems dominated by discs. On the other hand, non-barred galaxies tend to reach a rotationally-supported state at a slightly later stage, around redshift $z\approx 1-1.5$. While several factors can influence this process, our findings indicate that a higher angular momentum of gas at redshifts $z\approx 3-5$ favours the early formation of discs in barred galaxies.

\item[$\bullet$] Analyses of the TNG50 sample suggests that the MW has a strong bar~($0.35<A_2<0.6$) and it acquired a rotationally-supported disc at $z\approx 2 $. At that time, the mass of the disc component was about $0.5-1.5 \times 10^{9}$~\Msun, where the range of this estimate depends on the adopted MW stellar mass at $z=0$. 

\end{itemize}

Our analysis of TNG50, indicating a diminishing fraction of barred galaxies due to their late disc assembly, aligns with the ongoing challenge of bar formation in a cosmological framework. In the search for MW analogues in large-scale cosmological simulations, it is common to rely on the selection of present-day properties, such as colour, star formation rate, and mass. However, our results indicate that it is important to consider the influence of a galaxy's past in order to obtain more tailored MW analogues. We, therefore, suggest that it is worth incorporating star formation histories as a criterion for selecting MW analogues. In this case, the selection process would favour old massive discs, which are more likely to host strong bars at redshift $z=0$. This approach considers galaxies' evolutionary path and provides a more nuanced perspective on their formation and subsequent features, such as bars.

\section*{Acknowledgements}
Authors wish to thank the Illustris TNG collaboration for making their simulations publicly available. \newline

\section*{Data Availability}
Illustris TNG simulations are publicly available~\citep{2019ComAC...6....2N}\footnote{\url{https://www.tng-project.org}}.

\bibliographystyle{mnras}
\bibliography{references}

\begin{thebibliography}{}
\makeatletter
\relax
\def\mn@urlcharsother{\let\do\@makeother \do\$\do\&\do\#\do\^\do\_\do\%\do\~}
\def\mn@doi{\begingroup\mn@urlcharsother \@ifnextchar [ {\mn@doi@}
  {\mn@doi@[]}}
\def\mn@doi@[#1]#2{\def\@tempa{#1}\ifx\@tempa\@empty \href
  {http://dx.doi.org/#2} {doi:#2}\else \href {http://dx.doi.org/#2} {#1}\fi
  \endgroup}
\def\mn@eprint#1#2{\mn@eprint@#1:#2::\@nil}
\def\mn@eprint@arXiv#1{\href {http://arxiv.org/abs/#1} {{\tt arXiv:#1}}}
\def\mn@eprint@dblp#1{\href {http://dblp.uni-trier.de/rec/bibtex/#1.xml}
  {dblp:#1}}
\def\mn@eprint@#1:#2:#3:#4\@nil{\def\@tempa {#1}\def\@tempb {#2}\def\@tempc
  {#3}\ifx \@tempc \@empty \let \@tempc \@tempb \let \@tempb \@tempa \fi \ifx
  \@tempb \@empty \def\@tempb {arXiv}\fi \@ifundefined
  {mn@eprint@\@tempb}{\@tempb:\@tempc}{\expandafter \expandafter \csname
  mn@eprint@\@tempb\endcsname \expandafter{\@tempc}}}

\bibitem[\protect\citeauthoryear{{Abadi}, {Navarro}, {Steinmetz}  \&
  {Eke}}{{Abadi} et~al.}{2003a}]{2003ApJ...591..499A}
{Abadi} M.~G.,  {Navarro} J.~F.,  {Steinmetz} M.,   {Eke} V.~R.,  2003a,
  \mn@doi [\apj] {10.1086/375512}, \href
  {https://ui.adsabs.harvard.edu/abs/2003ApJ...591..499A} {591, 499}

\bibitem[\protect\citeauthoryear{{Abadi}, {Navarro}, {Steinmetz}  \&
  {Eke}}{{Abadi} et~al.}{2003b}]{2003ApJ...597...21A}
{Abadi} M.~G.,  {Navarro} J.~F.,  {Steinmetz} M.,   {Eke} V.~R.,  2003b,
  \mn@doi [\apj] {10.1086/378316}, \href
  {https://ui.adsabs.harvard.edu/abs/2003ApJ...597...21A} {597, 21}

\bibitem[\protect\citeauthoryear{{Agertz} et~al.,}{{Agertz}
  et~al.}{2021}]{2021MNRAS.503.5826A}
{Agertz} O.,  et~al., 2021, \mn@doi [\mnras] {10.1093/mnras/stab322}, \href
  {https://ui.adsabs.harvard.edu/abs/2021MNRAS.503.5826A} {503, 5826}

\bibitem[\protect\citeauthoryear{{Algorry} et~al.,}{{Algorry}
  et~al.}{2017}]{2017MNRAS.469.1054A}
{Algorry} D.~G.,  et~al., 2017, \mn@doi [\mnras] {10.1093/mnras/stx1008}, \href
  {https://ui.adsabs.harvard.edu/abs/2017MNRAS.469.1054A} {469, 1054}

\bibitem[\protect\citeauthoryear{{Anderson} et~al.,}{{Anderson}
  et~al.}{2023}]{2023arXiv230212788A}
{Anderson} S.~R.,  et~al., 2023, \mn@doi [arXiv e-prints]
  {10.48550/arXiv.2302.12788}, \href
  {https://ui.adsabs.harvard.edu/abs/2023arXiv230212788A} {p. arXiv:2302.12788}

\bibitem[\protect\citeauthoryear{{Athanassoula}}{{Athanassoula}}{1992}]{1992MNRAS.259..345A}
{Athanassoula} E.,  1992, \mn@doi [\mnras] {10.1093/mnras/259.2.345}, \href
  {https://ui.adsabs.harvard.edu/abs/1992MNRAS.259..345A} {259, 345}

\bibitem[\protect\citeauthoryear{{Athanassoula} \& {Misiriotis}}{{Athanassoula}
  \& {Misiriotis}}{2002}]{2002MNRAS.330...35A}
{Athanassoula} E.,  {Misiriotis} A.,  2002, \mn@doi [\mnras]
  {10.1046/j.1365-8711.2002.05028.x}, \href
  {https://ui.adsabs.harvard.edu/abs/2002MNRAS.330...35A} {330, 35}

\bibitem[\protect\citeauthoryear{{Athanassoula}, {Machado}  \&
  {Rodionov}}{{Athanassoula} et~al.}{2013}]{2013MNRAS.429.1949A}
{Athanassoula} E.,  {Machado} R. E.~G.,   {Rodionov} S.~A.,  2013, \mn@doi
  [\mnras] {10.1093/mnras/sts452}, \href
  {https://ui.adsabs.harvard.edu/abs/2013MNRAS.429.1949A} {429, 1949}

\bibitem[\protect\citeauthoryear{{Baba} \& {Kawata}}{{Baba} \&
  {Kawata}}{2020}]{2020MNRAS.492.4500B}
{Baba} J.,  {Kawata} D.,  2020, \mn@doi [\mnras] {10.1093/mnras/staa140}, \href
  {https://ui.adsabs.harvard.edu/abs/2020MNRAS.492.4500B} {492, 4500}

\bibitem[\protect\citeauthoryear{{Bardeen}}{{Bardeen}}{1975}]{1975IAUS...69..297B}
{Bardeen} J.~M.,  1975, in {Hayli} A.,  ed., ~ Vol. 69, Dynamics of the Solar
  Systems. p.~297

\bibitem[\protect\citeauthoryear{{Belokurov} \& {Kravtsov}}{{Belokurov} \&
  {Kravtsov}}{2022}]{2022MNRAS.514..689B}
{Belokurov} V.,  {Kravtsov} A.,  2022, \mn@doi [\mnras]
  {10.1093/mnras/stac1267}, \href
  {https://ui.adsabs.harvard.edu/abs/2022MNRAS.514..689B} {514, 689}

\bibitem[\protect\citeauthoryear{{Belokurov}, {Erkal}, {Evans}, {Koposov}  \&
  {Deason}}{{Belokurov} et~al.}{2018}]{2018MNRAS.478..611B}
{Belokurov} V.,  {Erkal} D.,  {Evans} N.~W.,  {Koposov} S.~E.,   {Deason}
  A.~J.,  2018, \mn@doi [\mnras] {10.1093/mnras/sty982}, \href
  {https://ui.adsabs.harvard.edu/abs/2018MNRAS.478..611B} {478, 611}

\bibitem[\protect\citeauthoryear{{Belokurov}, {Sanders}, {Fattahi}, {Smith},
  {Deason}, {Evans}  \& {Grand}}{{Belokurov}
  et~al.}{2020}]{2020MNRAS.494.3880B}
{Belokurov} V.,  {Sanders} J.~L.,  {Fattahi} A.,  {Smith} M.~C.,  {Deason}
  A.~J.,  {Evans} N.~W.,   {Grand} R. J.~J.,  2020, \mn@doi [\mnras]
  {10.1093/mnras/staa876}, \href
  {https://ui.adsabs.harvard.edu/abs/2020MNRAS.494.3880B} {494, 3880}

\bibitem[\protect\citeauthoryear{{Bensby}, {Feltzing}  \& {Oey}}{{Bensby}
  et~al.}{2014}]{2014A&A...562A..71B}
{Bensby} T.,  {Feltzing} S.,   {Oey} M.~S.,  2014, \mn@doi [\aap]
  {10.1051/0004-6361/201322631}, \href
  {https://ui.adsabs.harvard.edu/abs/2014A&A...562A..71B} {562, A71}

\bibitem[\protect\citeauthoryear{{Berentzen}, {Athanassoula}, {Heller}  \&
  {Fricke}}{{Berentzen} et~al.}{2004}]{2004MNRAS.347..220B}
{Berentzen} I.,  {Athanassoula} E.,  {Heller} C.~H.,   {Fricke} K.~J.,  2004,
  \mn@doi [\mnras] {10.1111/j.1365-2966.2004.07198.x}, \href
  {https://ui.adsabs.harvard.edu/abs/2004MNRAS.347..220B} {347, 220}

\bibitem[\protect\citeauthoryear{{Berentzen}, {Shlosman}, {Martinez-Valpuesta}
  \& {Heller}}{{Berentzen} et~al.}{2007}]{2007ApJ...666..189B}
{Berentzen} I.,  {Shlosman} I.,  {Martinez-Valpuesta} I.,   {Heller} C.~H.,
  2007, \mn@doi [\apj] {10.1086/520531}, \href
  {https://ui.adsabs.harvard.edu/abs/2007ApJ...666..189B} {666, 189}

\bibitem[\protect\citeauthoryear{{Bird}, {Kazantzidis}, {Weinberg}, {Guedes},
  {Callegari}, {Mayer}  \& {Madau}}{{Bird} et~al.}{2013}]{2013ApJ...773...43B}
{Bird} J.~C.,  {Kazantzidis} S.,  {Weinberg} D.~H.,  {Guedes} J.,  {Callegari}
  S.,  {Mayer} L.,   {Madau} P.,  2013, \mn@doi [\apj]
  {10.1088/0004-637X/773/1/43}, \href
  {https://ui.adsabs.harvard.edu/abs/2013ApJ...773...43B} {773, 43}

\bibitem[\protect\citeauthoryear{{Blitz} \& {Spergel}}{{Blitz} \&
  {Spergel}}{1991}]{1991ApJ...379..631B}
{Blitz} L.,  {Spergel} D.~N.,  1991, \mn@doi [\apj] {10.1086/170535}, \href
  {https://ui.adsabs.harvard.edu/abs/1991ApJ...379..631B} {379, 631}

\bibitem[\protect\citeauthoryear{{Bournaud} \& {Combes}}{{Bournaud} \&
  {Combes}}{2002}]{2002A&A...392...83B}
{Bournaud} F.,  {Combes} F.,  2002, \mn@doi [\aap]
  {10.1051/0004-6361:20020920}, \href
  {https://ui.adsabs.harvard.edu/abs/2002A&A...392...83B} {392, 83}

\bibitem[\protect\citeauthoryear{{Bournaud}, {Combes}  \& {Semelin}}{{Bournaud}
  et~al.}{2005}]{2005MNRAS.364L..18B}
{Bournaud} F.,  {Combes} F.,   {Semelin} B.,  2005, \mn@doi [\mnras]
  {10.1111/j.1745-3933.2005.00096.x}, \href
  {https://ui.adsabs.harvard.edu/abs/2005MNRAS.364L..18B} {364, L18}

\bibitem[\protect\citeauthoryear{{Bovy}, {Leung}, {Hunt}, {Mackereth},
  {Garc{\'\i}a-Hern{\'a}ndez}  \& {Roman-Lopes}}{{Bovy}
  et~al.}{2019}]{2019MNRAS.490.4740B}
{Bovy} J.,  {Leung} H.~W.,  {Hunt} J. A.~S.,  {Mackereth} J.~T.,
  {Garc{\'\i}a-Hern{\'a}ndez} D.~A.,   {Roman-Lopes} A.,  2019, \mn@doi
  [\mnras] {10.1093/mnras/stz2891}, \href
  {https://ui.adsabs.harvard.edu/abs/2019MNRAS.490.4740B} {490, 4740}

\bibitem[\protect\citeauthoryear{{Brook}, {Kawata}, {Gibson}  \&
  {Freeman}}{{Brook} et~al.}{2004}]{brook04}
{Brook} C.~B.,  {Kawata} D.,  {Gibson} B.~K.,   {Freeman} K.~C.,  2004, \mn@doi
  [\apj] {10.1086/422709}, \href
  {https://ui.adsabs.harvard.edu/abs/2004ApJ...612..894B} {612, 894}

\bibitem[\protect\citeauthoryear{{Buck}}{{Buck}}{2020}]{2020MNRAS.491.5435B}
{Buck} T.,  2020, \mn@doi [\mnras] {10.1093/mnras/stz3289}, \href
  {https://ui.adsabs.harvard.edu/abs/2020MNRAS.491.5435B} {491, 5435}

\bibitem[\protect\citeauthoryear{{Buck}, {Ness}, {Macci{\`o}}, {Obreja}  \&
  {Dutton}}{{Buck} et~al.}{2018}]{2018ApJ...861...88B}
{Buck} T.,  {Ness} M.~K.,  {Macci{\`o}} A.~V.,  {Obreja} A.,   {Dutton} A.~A.,
  2018, \mn@doi [\apj] {10.3847/1538-4357/aac890}, \href
  {https://ui.adsabs.harvard.edu/abs/2018ApJ...861...88B} {861, 88}

\bibitem[\protect\citeauthoryear{{Burkert}, {Truran}  \& {Hensler}}{{Burkert}
  et~al.}{1992}]{1992ApJ...391..651B}
{Burkert} A.,  {Truran} J.~W.,   {Hensler} G.,  1992, \mn@doi [\apj]
  {10.1086/171378}, \href
  {https://ui.adsabs.harvard.edu/abs/1992ApJ...391..651B} {391, 651}

\bibitem[\protect\citeauthoryear{{Callingham} et~al.,}{{Callingham}
  et~al.}{2019}]{2019MNRAS.484.5453C}
{Callingham} T.~M.,  et~al., 2019, \mn@doi [\mnras] {10.1093/mnras/stz365},
  \href {https://ui.adsabs.harvard.edu/abs/2019MNRAS.484.5453C} {484, 5453}

\bibitem[\protect\citeauthoryear{{Cao}, {Mao}, {Nataf}, {Rattenbury}  \&
  {Gould}}{{Cao} et~al.}{2013}]{2013MNRAS.434..595C}
{Cao} L.,  {Mao} S.,  {Nataf} D.,  {Rattenbury} N.~J.,   {Gould} A.,  2013,
  \mn@doi [\mnras] {10.1093/mnras/stt1045}, \href
  {https://ui.adsabs.harvard.edu/abs/2013MNRAS.434..595C} {434, 595}

\bibitem[\protect\citeauthoryear{{Cautun} et~al.,}{{Cautun}
  et~al.}{2020}]{2020MNRAS.494.4291C}
{Cautun} M.,  et~al., 2020, \mn@doi [\mnras] {10.1093/mnras/staa1017}, \href
  {https://ui.adsabs.harvard.edu/abs/2020MNRAS.494.4291C} {494, 4291}

\bibitem[\protect\citeauthoryear{{Chiappini}, {Matteucci}, {Beers}  \&
  {Nomoto}}{{Chiappini} et~al.}{1999}]{1999ApJ...515..226C}
{Chiappini} C.,  {Matteucci} F.,  {Beers} T.~C.,   {Nomoto} K.,  1999, \mn@doi
  [\apj] {10.1086/307006}, \href
  {https://ui.adsabs.harvard.edu/abs/1999ApJ...515..226C} {515, 226}

\bibitem[\protect\citeauthoryear{{Clarke} et~al.,}{{Clarke}
  et~al.}{2019}]{2019MNRAS.484.3476C}
{Clarke} A.~J.,  et~al., 2019, \mn@doi [\mnras] {10.1093/mnras/stz104}, \href
  {https://ui.adsabs.harvard.edu/abs/2019MNRAS.484.3476C} {484, 3476}

\bibitem[\protect\citeauthoryear{{Cole}, {Debattista}, {Erwin}, {Earp}  \&
  {Ro{\v{s}}kar}}{{Cole} et~al.}{2014}]{2014MNRAS.445.3352C}
{Cole} D.~R.,  {Debattista} V.~P.,  {Erwin} P.,  {Earp} S. W.~F.,
  {Ro{\v{s}}kar} R.,  2014, \mn@doi [\mnras] {10.1093/mnras/stu1985}, \href
  {https://ui.adsabs.harvard.edu/abs/2014MNRAS.445.3352C} {445, 3352}

\bibitem[\protect\citeauthoryear{{Combes} \& {Sanders}}{{Combes} \&
  {Sanders}}{1981}]{1981A&A....96..164C}
{Combes} F.,  {Sanders} R.~H.,  1981, \aap, \href
  {https://ui.adsabs.harvard.edu/abs/1981A&A....96..164C} {96, 164}

\bibitem[\protect\citeauthoryear{{Conroy} et~al.,}{{Conroy}
  et~al.}{2022}]{2022arXiv220402989C}
{Conroy} C.,  et~al., 2022, \mn@doi [arXiv e-prints]
  {10.48550/arXiv.2204.02989}, \href
  {https://ui.adsabs.harvard.edu/abs/2022arXiv220402989C} {p. arXiv:2204.02989}

\bibitem[\protect\citeauthoryear{{Correa Magnus} \& {Vasiliev}}{{Correa Magnus}
  \& {Vasiliev}}{2022}]{2022MNRAS.511.2610C}
{Correa Magnus} L.,  {Vasiliev} E.,  2022, \mn@doi [\mnras]
  {10.1093/mnras/stab3726}, \href
  {https://ui.adsabs.harvard.edu/abs/2022MNRAS.511.2610C} {511, 2610}

\bibitem[\protect\citeauthoryear{{Cortese} et~al.,}{{Cortese}
  et~al.}{2016}]{2016MNRAS.463..170C}
{Cortese} L.,  et~al., 2016, \mn@doi [\mnras] {10.1093/mnras/stw1891}, \href
  {https://ui.adsabs.harvard.edu/abs/2016MNRAS.463..170C} {463, 170}

\bibitem[\protect\citeauthoryear{{Curtis}}{{Curtis}}{1918}]{1918PLicO..13....9C}
{Curtis} H.~D.,  1918, Publications of Lick Observatory, \href
  {https://ui.adsabs.harvard.edu/abs/1918PLicO..13....9C} {13, 9}

\bibitem[\protect\citeauthoryear{{Danovich}, {Dekel}, {Hahn}  \&
  {Teyssier}}{{Danovich} et~al.}{2012}]{2012MNRAS.422.1732D}
{Danovich} M.,  {Dekel} A.,  {Hahn} O.,   {Teyssier} R.,  2012, \mn@doi
  [\mnras] {10.1111/j.1365-2966.2012.20751.x}, \href
  {https://ui.adsabs.harvard.edu/abs/2012MNRAS.422.1732D} {422, 1732}

\bibitem[\protect\citeauthoryear{{Danovich}, {Dekel}, {Hahn}, {Ceverino}  \&
  {Primack}}{{Danovich} et~al.}{2015}]{2015MNRAS.449.2087D}
{Danovich} M.,  {Dekel} A.,  {Hahn} O.,  {Ceverino} D.,   {Primack} J.,  2015,
  \mn@doi [\mnras] {10.1093/mnras/stv270}, \href
  {https://ui.adsabs.harvard.edu/abs/2015MNRAS.449.2087D} {449, 2087}

\bibitem[\protect\citeauthoryear{{Dantas} et~al.,}{{Dantas}
  et~al.}{2023}]{2023A&A...669A..96D}
{Dantas} M.~L.~L.,  et~al., 2023, \mn@doi [\aap] {10.1051/0004-6361/202243667},
  \href {https://ui.adsabs.harvard.edu/abs/2023A&A...669A..96D} {669, A96}

\bibitem[\protect\citeauthoryear{{Deason}, {Fattahi}, {Belokurov}, {Evans},
  {Grand}, {Marinacci}  \& {Pakmor}}{{Deason}
  et~al.}{2019}]{2019MNRAS.485.3514D}
{Deason} A.~J.,  {Fattahi} A.,  {Belokurov} V.,  {Evans} N.~W.,  {Grand} R.
  J.~J.,  {Marinacci} F.,   {Pakmor} R.,  2019, \mn@doi [\mnras]
  {10.1093/mnras/stz623}, \href
  {https://ui.adsabs.harvard.edu/abs/2019MNRAS.485.3514D} {485, 3514}

\bibitem[\protect\citeauthoryear{{Deason} et~al.,}{{Deason}
  et~al.}{2021}]{2021MNRAS.501.5964D}
{Deason} A.~J.,  et~al., 2021, \mn@doi [\mnras] {10.1093/mnras/staa3984}, \href
  {https://ui.adsabs.harvard.edu/abs/2021MNRAS.501.5964D} {501, 5964}

\bibitem[\protect\citeauthoryear{{Debattista}, {Gonzalez}, {Sanderson},
  {El-Badry}, {Garrison-Kimmel}, {Wetzel}, {Faucher-Gigu{\`e}re}  \&
  {Hopkins}}{{Debattista} et~al.}{2019}]{2019MNRAS.485.5073D}
{Debattista} V.~P.,  {Gonzalez} O.~A.,  {Sanderson} R.~E.,  {El-Badry} K.,
  {Garrison-Kimmel} S.,  {Wetzel} A.,  {Faucher-Gigu{\`e}re} C.-A.,   {Hopkins}
  P.~F.,  2019, \mn@doi [\mnras] {10.1093/mnras/stz746}, \href
  {https://ui.adsabs.harvard.edu/abs/2019MNRAS.485.5073D} {485, 5073}

\bibitem[\protect\citeauthoryear{{Di Matteo}, {Haywood}, {Lehnert}, {Katz},
  {Khoperskov}, {Snaith}, {G{\'o}mez}  \& {Robichon}}{{Di Matteo}
  et~al.}{2019}]{2019A&A...632A...4D}
{Di Matteo} P.,  {Haywood} M.,  {Lehnert} M.~D.,  {Katz} D.,  {Khoperskov} S.,
  {Snaith} O.~N.,  {G{\'o}mez} A.,   {Robichon} N.,  2019, \mn@doi [\aap]
  {10.1051/0004-6361/201834929}, \href
  {https://ui.adsabs.harvard.edu/abs/2019A&A...632A...4D} {632, A4}

\bibitem[\protect\citeauthoryear{{Di Matteo}, {Spite}, {Haywood}, {Bonifacio},
  {G{\'o}mez}, {Spite}  \& {Caffau}}{{Di Matteo}
  et~al.}{2020}]{2020A&A...636A.115D}
{Di Matteo} P.,  {Spite} M.,  {Haywood} M.,  {Bonifacio} P.,  {G{\'o}mez} A.,
  {Spite} F.,   {Caffau} E.,  2020, \mn@doi [\aap]
  {10.1051/0004-6361/201937016}, \href
  {https://ui.adsabs.harvard.edu/abs/2020A&A...636A.115D} {636, A115}

\bibitem[\protect\citeauthoryear{{Dubois} et~al.,}{{Dubois}
  et~al.}{2014}]{2014MNRAS.444.1453D}
{Dubois} Y.,  et~al., 2014, \mn@doi [\mnras] {10.1093/mnras/stu1227}, \href
  {https://ui.adsabs.harvard.edu/abs/2014MNRAS.444.1453D} {444, 1453}

\bibitem[\protect\citeauthoryear{{Dwek} et~al.,}{{Dwek}
  et~al.}{1995}]{1995ApJ...445..716D}
{Dwek} E.,  et~al., 1995, \mn@doi [\apj] {10.1086/175734}, \href
  {https://ui.adsabs.harvard.edu/abs/1995ApJ...445..716D} {445, 716}

\bibitem[\protect\citeauthoryear{{Efstathiou}, {Lake}  \&
  {Negroponte}}{{Efstathiou} et~al.}{1982}]{1982MNRAS.199.1069E}
{Efstathiou} G.,  {Lake} G.,   {Negroponte} J.,  1982, \mn@doi [\mnras]
  {10.1093/mnras/199.4.1069}, \href
  {https://ui.adsabs.harvard.edu/abs/1982MNRAS.199.1069E} {199, 1069}

\bibitem[\protect\citeauthoryear{{Erkal} et~al.,}{{Erkal}
  et~al.}{2019}]{2019MNRAS.487.2685E}
{Erkal} D.,  et~al., 2019, \mn@doi [\mnras] {10.1093/mnras/stz1371}, \href
  {https://ui.adsabs.harvard.edu/abs/2019MNRAS.487.2685E} {487, 2685}

\bibitem[\protect\citeauthoryear{{Erkal}, {Belokurov}  \& {Parkin}}{{Erkal}
  et~al.}{2020}]{2020MNRAS.498.5574E}
{Erkal} D.,  {Belokurov} V.~A.,   {Parkin} D.~L.,  2020, \mn@doi [\mnras]
  {10.1093/mnras/staa2840}, \href
  {https://ui.adsabs.harvard.edu/abs/2020MNRAS.498.5574E} {498, 5574}

\bibitem[\protect\citeauthoryear{{Eskridge} et~al.,}{{Eskridge}
  et~al.}{2000}]{2000AJ....119..536E}
{Eskridge} P.~B.,  et~al., 2000, \mn@doi [\aj] {10.1086/301203}, \href
  {https://ui.adsabs.harvard.edu/abs/2000AJ....119..536E} {119, 536}

\bibitem[\protect\citeauthoryear{{Fall} \& {Romanowsky}}{{Fall} \&
  {Romanowsky}}{2018}]{2018ApJ...868..133F}
{Fall} S.~M.,  {Romanowsky} A.~J.,  2018, \mn@doi [\apj]
  {10.3847/1538-4357/aaeb27}, \href
  {https://ui.adsabs.harvard.edu/abs/2018ApJ...868..133F} {868, 133}

\bibitem[\protect\citeauthoryear{{Fragkoudi}, {Grand}, {Pakmor}, {Springel},
  {White}, {Marinacci}, {Gomez}  \& {Navarro}}{{Fragkoudi}
  et~al.}{2021}]{2021A&A...650L..16F}
{Fragkoudi} F.,  {Grand} R.~J.~J.,  {Pakmor} R.,  {Springel} V.,  {White}
  S.~D.~M.,  {Marinacci} F.,  {Gomez} F.~A.,   {Navarro} J.~F.,  2021, \mn@doi
  [\aap] {10.1051/0004-6361/202140320}, \href
  {https://ui.adsabs.harvard.edu/abs/2021A&A...650L..16F} {650, L16}

\bibitem[\protect\citeauthoryear{{Frankel} et~al.,}{{Frankel}
  et~al.}{2022}]{2022ApJ...940...61F}
{Frankel} N.,  et~al., 2022, \mn@doi [\apj] {10.3847/1538-4357/ac9972}, \href
  {https://ui.adsabs.harvard.edu/abs/2022ApJ...940...61F} {940, 61}

\bibitem[\protect\citeauthoryear{{Fraser-McKelvie} et~al.,}{{Fraser-McKelvie}
  et~al.}{2020}]{2020MNRAS.495.4158F}
{Fraser-McKelvie} A.,  et~al., 2020, \mn@doi [\mnras] {10.1093/mnras/staa1416},
  \href {https://ui.adsabs.harvard.edu/abs/2020MNRAS.495.4158F} {495, 4158}

\bibitem[\protect\citeauthoryear{{Fraternali}, {Karim}, {Magnelli},
  {G{\'o}mez-Guijarro}, {Jim{\'e}nez-Andrade}  \& {Posses}}{{Fraternali}
  et~al.}{2021}]{2021A&A...647A.194F}
{Fraternali} F.,  {Karim} A.,  {Magnelli} B.,  {G{\'o}mez-Guijarro} C.,
  {Jim{\'e}nez-Andrade} E.~F.,   {Posses} A.~C.,  2021, \mn@doi [\aap]
  {10.1051/0004-6361/202039807}, \href
  {https://ui.adsabs.harvard.edu/abs/2021A&A...647A.194F} {647, A194}

\bibitem[\protect\citeauthoryear{{Freeman}}{{Freeman}}{1996}]{1996ASPC...91....1F}
{Freeman} K.~C.,  1996, in {Buta} R.,  {Crocker} D.~A.,   {Elmegreen} B.~G.,
  eds,  Astronomical Society of the Pacific Conference Series Vol. 91, IAU
  Colloq. 157: Barred Galaxies. p.~1

\bibitem[\protect\citeauthoryear{{Gadotti}, {Seidel},
  {S{\'a}nchez-Bl{\'a}zquez}, {Falc{\'o}n-Barroso}, {Husemann}, {Coelho}  \&
  {P{\'e}rez}}{{Gadotti} et~al.}{2015}]{2015A&A...584A..90G}
{Gadotti} D.~A.,  {Seidel} M.~K.,  {S{\'a}nchez-Bl{\'a}zquez} P.,
  {Falc{\'o}n-Barroso} J.,  {Husemann} B.,  {Coelho} P.,   {P{\'e}rez} I.,
  2015, \mn@doi [\aap] {10.1051/0004-6361/201526677}, \href
  {https://ui.adsabs.harvard.edu/abs/2015A&A...584A..90G} {584, A90}

\bibitem[\protect\citeauthoryear{{Gaia Collaboration} et~al.,}{{Gaia
  Collaboration} et~al.}{2023}]{2023A&A...674A..37G}
{Gaia Collaboration} et~al., 2023, \mn@doi [\aap]
  {10.1051/0004-6361/202243797}, \href
  {https://ui.adsabs.harvard.edu/abs/2023A&A...674A..37G} {674, A37}

\bibitem[\protect\citeauthoryear{{Gargiulo} et~al.,}{{Gargiulo}
  et~al.}{2022}]{2022MNRAS.512.2537G}
{Gargiulo} I.~D.,  et~al., 2022, \mn@doi [\mnras] {10.1093/mnras/stac629},
  \href {https://ui.adsabs.harvard.edu/abs/2022MNRAS.512.2537G} {512, 2537}

\bibitem[\protect\citeauthoryear{{Genel}, {Fall}, {Hernquist}, {Vogelsberger},
  {Snyder}, {Rodriguez-Gomez}, {Sijacki}  \& {Springel}}{{Genel}
  et~al.}{2015}]{2015ApJ...804L..40G}
{Genel} S.,  {Fall} S.~M.,  {Hernquist} L.,  {Vogelsberger} M.,  {Snyder}
  G.~F.,  {Rodriguez-Gomez} V.,  {Sijacki} D.,   {Springel} V.,  2015, \mn@doi
  [\apjl] {10.1088/2041-8205/804/2/L40}, \href
  {https://ui.adsabs.harvard.edu/abs/2015ApJ...804L..40G} {804, L40}

\bibitem[\protect\citeauthoryear{{Ghosh}}{{Ghosh}}{2023}]{Soumavo}
{Ghosh} et~al. i.~p.,  2023, \mnras

\bibitem[\protect\citeauthoryear{{Goldreich} \& {Lynden-Bell}}{{Goldreich} \&
  {Lynden-Bell}}{1965}]{1965MNRAS.130..125G}
{Goldreich} P.,  {Lynden-Bell} D.,  1965, \mn@doi [\mnras]
  {10.1093/mnras/130.2.125}, \href
  {https://ui.adsabs.harvard.edu/abs/1965MNRAS.130..125G} {130, 125}

\bibitem[\protect\citeauthoryear{{Gonz{\'a}lez Delgado} et~al.,}{{Gonz{\'a}lez
  Delgado} et~al.}{2017}]{2017A&A...607A.128G}
{Gonz{\'a}lez Delgado} R.~M.,  et~al., 2017, \mn@doi [\aap]
  {10.1051/0004-6361/201730883}, \href
  {https://ui.adsabs.harvard.edu/abs/2017A&A...607A.128G} {607, A128}

\bibitem[\protect\citeauthoryear{{Governato}, {Willman}, {Mayer}, {Brooks},
  {Stinson}, {Valenzuela}, {Wadsley}  \& {Quinn}}{{Governato}
  et~al.}{2007}]{2007MNRAS.374.1479G}
{Governato} F.,  {Willman} B.,  {Mayer} L.,  {Brooks} A.,  {Stinson} G.,
  {Valenzuela} O.,  {Wadsley} J.,   {Quinn} T.,  2007, \mn@doi [\mnras]
  {10.1111/j.1365-2966.2006.11266.x}, \href
  {https://ui.adsabs.harvard.edu/abs/2007MNRAS.374.1479G} {374, 1479}

\bibitem[\protect\citeauthoryear{{Grand} et~al.,}{{Grand}
  et~al.}{2017}]{2017MNRAS.467..179G}
{Grand} R. J.~J.,  et~al., 2017, \mn@doi [\mnras] {10.1093/mnras/stx071}, \href
  {https://ui.adsabs.harvard.edu/abs/2017MNRAS.467..179G} {467, 179}

\bibitem[\protect\citeauthoryear{{Grand} et~al.,}{{Grand}
  et~al.}{2018}]{2018MNRAS.474.3629G}
{Grand} R. J.~J.,  et~al., 2018, \mn@doi [\mnras] {10.1093/mnras/stx3025},
  \href {https://ui.adsabs.harvard.edu/abs/2018MNRAS.474.3629G} {474, 3629}

\bibitem[\protect\citeauthoryear{{Gurvich} et~al.,}{{Gurvich}
  et~al.}{2023}]{2023MNRAS.519.2598G}
{Gurvich} A.~B.,  et~al., 2023, \mn@doi [\mnras] {10.1093/mnras/stac3712},
  \href {https://ui.adsabs.harvard.edu/abs/2023MNRAS.519.2598G} {519, 2598}

\bibitem[\protect\citeauthoryear{{Haywood}, {Di Matteo}, {Lehnert}, {Katz}  \&
  {G{\'o}mez}}{{Haywood} et~al.}{2013}]{2013A&A...560A.109H}
{Haywood} M.,  {Di Matteo} P.,  {Lehnert} M.~D.,  {Katz} D.,   {G{\'o}mez} A.,
  2013, \mn@doi [\aap] {10.1051/0004-6361/201321397}, \href
  {https://ui.adsabs.harvard.edu/abs/2013A&A...560A.109H} {560, A109}

\bibitem[\protect\citeauthoryear{{Haywood}, {Di Matteo}, {Snaith}  \&
  {Lehnert}}{{Haywood} et~al.}{2015}]{2015A&A...579A...5H}
{Haywood} M.,  {Di Matteo} P.,  {Snaith} O.,   {Lehnert} M.~D.,  2015, \mn@doi
  [\aap] {10.1051/0004-6361/201425459}, \href
  {https://ui.adsabs.harvard.edu/abs/2015A&A...579A...5H} {579, A5}

\bibitem[\protect\citeauthoryear{{Haywood}, {Di Matteo}, {Lehnert}, {Snaith},
  {Fragkoudi}  \& {Khoperskov}}{{Haywood} et~al.}{2018a}]{2018A&A...618A..78H}
{Haywood} M.,  {Di Matteo} P.,  {Lehnert} M.,  {Snaith} O.,  {Fragkoudi} F.,
  {Khoperskov} S.,  2018a, \mn@doi [\aap] {10.1051/0004-6361/201731363}, \href
  {https://ui.adsabs.harvard.edu/abs/2018A&A...618A..78H} {618, A78}

\bibitem[\protect\citeauthoryear{{Haywood}, {Di Matteo}, {Lehnert}, {Snaith},
  {Khoperskov}  \& {G{\'o}mez}}{{Haywood} et~al.}{2018b}]{2018ApJ...863..113H}
{Haywood} M.,  {Di Matteo} P.,  {Lehnert} M.~D.,  {Snaith} O.,  {Khoperskov}
  S.,   {G{\'o}mez} A.,  2018b, \mn@doi [\apj] {10.3847/1538-4357/aad235},
  \href {https://ui.adsabs.harvard.edu/abs/2018ApJ...863..113H} {863, 113}

\bibitem[\protect\citeauthoryear{{Haywood}, {Khoperskov}, {Cerqui}, {Di
  Matteo}, {Katz}  \& {Snaith}}{{Haywood} et~al.}{2024}]{2024arXiv240308963H}
{Haywood} M.,  {Khoperskov} S.,  {Cerqui} V.,  {Di Matteo} P.,  {Katz} D.,
  {Snaith} O.,  2024, \mn@doi [arXiv e-prints] {10.48550/arXiv.2403.08963},
  \href {https://ui.adsabs.harvard.edu/abs/2024arXiv240308963H} {p.
  arXiv:2403.08963}

\bibitem[\protect\citeauthoryear{{Helmi}, {Babusiaux}, {Koppelman}, {Massari},
  {Veljanoski}  \& {Brown}}{{Helmi} et~al.}{2018}]{2018Natur.563...85H}
{Helmi} A.,  {Babusiaux} C.,  {Koppelman} H.~H.,  {Massari} D.,  {Veljanoski}
  J.,   {Brown} A. G.~A.,  2018, \mn@doi [\nat] {10.1038/s41586-018-0625-x},
  \href {https://ui.adsabs.harvard.edu/abs/2018Natur.563...85H} {563, 85}

\bibitem[\protect\citeauthoryear{{Hilmi} et~al.,}{{Hilmi}
  et~al.}{2020}]{2020MNRAS.497..933H}
{Hilmi} T.,  et~al., 2020, \mn@doi [\mnras] {10.1093/mnras/staa1934}, \href
  {https://ui.adsabs.harvard.edu/abs/2020MNRAS.497..933H} {497, 933}

\bibitem[\protect\citeauthoryear{{Hohl}}{{Hohl}}{1971}]{1971ApJ...168..343H}
{Hohl} F.,  1971, \mn@doi [\apj] {10.1086/151091}, \href
  {https://ui.adsabs.harvard.edu/abs/1971ApJ...168..343H} {168, 343}

\bibitem[\protect\citeauthoryear{{Hopkins} et~al.,}{{Hopkins}
  et~al.}{2023}]{2023MNRAS.tmp.1847H}
{Hopkins} P.~F.,  et~al., 2023, \mn@doi [\mnras] {10.1093/mnras/stad1902},
  \href {https://ui.adsabs.harvard.edu/abs/2023MNRAS.tmp.1847H} {}

\bibitem[\protect\citeauthoryear{{Hubble}}{{Hubble}}{1926}]{1926ApJ....64..321H}
{Hubble} E.~P.,  1926, \mn@doi [\apj] {10.1086/143018}, \href
  {https://ui.adsabs.harvard.edu/abs/1926ApJ....64..321H} {64, 321}

\bibitem[\protect\citeauthoryear{{Izquierdo-Villalba}
  et~al.,}{{Izquierdo-Villalba} et~al.}{2022}]{2022MNRAS.514.1006I}
{Izquierdo-Villalba} D.,  et~al., 2022, \mn@doi [\mnras]
  {10.1093/mnras/stac1413}, \href
  {https://ui.adsabs.harvard.edu/abs/2022MNRAS.514.1006I} {514, 1006}

\bibitem[\protect\citeauthoryear{{Johnson}}{{Johnson}}{1957}]{1957AJ.....62...19J}
{Johnson} H.~M.,  1957, \mn@doi [\aj] {10.1086/107441}, \href
  {https://ui.adsabs.harvard.edu/abs/1957AJ.....62...19J} {62, 19}

\bibitem[\protect\citeauthoryear{{Julian} \& {Toomre}}{{Julian} \&
  {Toomre}}{1966}]{1966ApJ...146..810J}
{Julian} W.~H.,  {Toomre} A.,  1966, \mn@doi [\apj] {10.1086/148957}, \href
  {https://ui.adsabs.harvard.edu/abs/1966ApJ...146..810J} {146, 810}

\bibitem[\protect\citeauthoryear{{Kalnajs}}{{Kalnajs}}{1972}]{1972ApJ...175...63K}
{Kalnajs} A.~J.,  1972, \mn@doi [\apj] {10.1086/151538}, \href
  {https://ui.adsabs.harvard.edu/abs/1972ApJ...175...63K} {175, 63}

\bibitem[\protect\citeauthoryear{{Kassin} et~al.,}{{Kassin}
  et~al.}{2012}]{2012ApJ...758..106K}
{Kassin} S.~A.,  et~al., 2012, \mn@doi [\apj] {10.1088/0004-637X/758/2/106},
  \href {https://ui.adsabs.harvard.edu/abs/2012ApJ...758..106K} {758, 106}

\bibitem[\protect\citeauthoryear{{Kataria} \& {Das}}{{Kataria} \&
  {Das}}{2018}]{2018MNRAS.475.1653K}
{Kataria} S.~K.,  {Das} M.,  2018, \mn@doi [\mnras] {10.1093/mnras/stx3279},
  \href {https://ui.adsabs.harvard.edu/abs/2018MNRAS.475.1653K} {475, 1653}

\bibitem[\protect\citeauthoryear{{Khoperskov}, {Haywood}, {Di Matteo},
  {Lehnert}  \& {Combes}}{{Khoperskov} et~al.}{2018}]{2018A&A...609A..60K}
{Khoperskov} S.,  {Haywood} M.,  {Di Matteo} P.,  {Lehnert} M.~D.,   {Combes}
  F.,  2018, \mn@doi [\aap] {10.1051/0004-6361/201731211}, \href
  {https://ui.adsabs.harvard.edu/abs/2018A&A...609A..60K} {609, A60}

\bibitem[\protect\citeauthoryear{{Khoperskov}, {Di Matteo}, {Haywood},
  {G{\'o}mez}  \& {Snaith}}{{Khoperskov} et~al.}{2020}]{2020A&A...638A.144K}
{Khoperskov} S.,  {Di Matteo} P.,  {Haywood} M.,  {G{\'o}mez} A.,   {Snaith}
  O.~N.,  2020, \mn@doi [\aap] {10.1051/0004-6361/201937188}, \href
  {https://ui.adsabs.harvard.edu/abs/2020A&A...638A.144K} {638, A144}

\bibitem[\protect\citeauthoryear{{Khoperskov}, {Haywood}, {Snaith}, {Di
  Matteo}, {Lehnert}, {Vasiliev}, {Naroenkov}  \& {Berczik}}{{Khoperskov}
  et~al.}{2021}]{2021MNRAS.501.5176K}
{Khoperskov} S.,  {Haywood} M.,  {Snaith} O.,  {Di Matteo} P.,  {Lehnert} M.,
  {Vasiliev} E.,  {Naroenkov} S.,   {Berczik} P.,  2021, \mn@doi [\mnras]
  {10.1093/mnras/staa3996}, \href
  {https://ui.adsabs.harvard.edu/abs/2021MNRAS.501.5176K} {501, 5176}

\bibitem[\protect\citeauthoryear{{Khoperskov} et~al.,}{{Khoperskov}
  et~al.}{2023}]{2023A&A...677A..91K}
{Khoperskov} S.,  et~al., 2023, \mn@doi [\aap] {10.1051/0004-6361/202244234},
  \href {https://ui.adsabs.harvard.edu/abs/2023A&A...677A..91K} {677, A91}

\bibitem[\protect\citeauthoryear{{Knapen}, {Shlosman}  \& {Peletier}}{{Knapen}
  et~al.}{2000}]{2000ApJ...529...93K}
{Knapen} J.~H.,  {Shlosman} I.,   {Peletier} R.~F.,  2000, \mn@doi [\apj]
  {10.1086/308266}, \href
  {https://ui.adsabs.harvard.edu/abs/2000ApJ...529...93K} {529, 93}

\bibitem[\protect\citeauthoryear{{Kraljic}, {Bournaud}  \& {Martig}}{{Kraljic}
  et~al.}{2012}]{2012ApJ...757...60K}
{Kraljic} K.,  {Bournaud} F.,   {Martig} M.,  2012, \mn@doi [\apj]
  {10.1088/0004-637X/757/1/60}, \href
  {https://ui.adsabs.harvard.edu/abs/2012ApJ...757...60K} {757, 60}

\bibitem[\protect\citeauthoryear{{Kretschmer}, {Agertz}  \&
  {Teyssier}}{{Kretschmer} et~al.}{2020}]{2020MNRAS.497.4346K}
{Kretschmer} M.,  {Agertz} O.,   {Teyssier} R.,  2020, \mn@doi [\mnras]
  {10.1093/mnras/staa2243}, \href
  {https://ui.adsabs.harvard.edu/abs/2020MNRAS.497.4346K} {497, 4346}

\bibitem[\protect\citeauthoryear{{Kretschmer}, {Dekel}  \&
  {Teyssier}}{{Kretschmer} et~al.}{2022}]{2022MNRAS.510.3266K}
{Kretschmer} M.,  {Dekel} A.,   {Teyssier} R.,  2022, \mn@doi [\mnras]
  {10.1093/mnras/stab3648}, \href
  {https://ui.adsabs.harvard.edu/abs/2022MNRAS.510.3266K} {510, 3266}

\bibitem[\protect\citeauthoryear{{Kruk} et~al.,}{{Kruk}
  et~al.}{2018}]{2018MNRAS.473.4731K}
{Kruk} S.~J.,  et~al., 2018, \mn@doi [\mnras] {10.1093/mnras/stx2605}, \href
  {https://ui.adsabs.harvard.edu/abs/2018MNRAS.473.4731K} {473, 4731}

\bibitem[\protect\citeauthoryear{{Lehnert}, {Di Matteo}, {Haywood}  \&
  {Snaith}}{{Lehnert} et~al.}{2014}]{2014ApJ...789L..30L}
{Lehnert} M.~D.,  {Di Matteo} P.,  {Haywood} M.,   {Snaith} O.~N.,  2014,
  \mn@doi [\apjl] {10.1088/2041-8205/789/2/L30}, \href
  {https://ui.adsabs.harvard.edu/abs/2014ApJ...789L..30L} {789, L30}

\bibitem[\protect\citeauthoryear{{Lelli}, {Di Teodoro}, {Fraternali}, {Man},
  {Zhang}, {De Breuck}, {Davis}  \& {Maiolino}}{{Lelli}
  et~al.}{2021}]{2021Sci...371..713L}
{Lelli} F.,  {Di Teodoro} E.~M.,  {Fraternali} F.,  {Man} A. W.~S.,  {Zhang}
  Z.-Y.,  {De Breuck} C.,  {Davis} T.~A.,   {Maiolino} R.,  2021, \mn@doi
  [Science] {10.1126/science.abc1893}, \href
  {https://ui.adsabs.harvard.edu/abs/2021Sci...371..713L} {371, 713}

\bibitem[\protect\citeauthoryear{{Libeskind}, {Hoffman}, {Forero-Romero},
  {Gottl{\"o}ber}, {Knebe}, {Steinmetz}  \& {Klypin}}{{Libeskind}
  et~al.}{2013}]{2013MNRAS.428.2489L}
{Libeskind} N.~I.,  {Hoffman} Y.,  {Forero-Romero} J.,  {Gottl{\"o}ber} S.,
  {Knebe} A.,  {Steinmetz} M.,   {Klypin} A.,  2013, \mn@doi [\mnras]
  {10.1093/mnras/sts216}, \href
  {https://ui.adsabs.harvard.edu/abs/2013MNRAS.428.2489L} {428, 2489}

\bibitem[\protect\citeauthoryear{{{\L}okas}}{{{\L}okas}}{2021}]{2021A&A...647A.143L}
{{\L}okas} E.~L.,  2021, \mn@doi [\aap] {10.1051/0004-6361/202040056}, \href
  {https://ui.adsabs.harvard.edu/abs/2021A&A...647A.143L} {647, A143}

\bibitem[\protect\citeauthoryear{{{\L}okas}}{{{\L}okas}}{2022}]{2022A&A...668L...3L}
{{\L}okas} E.~L.,  2022, \mn@doi [\aap] {10.1051/0004-6361/202245056}, \href
  {https://ui.adsabs.harvard.edu/abs/2022A&A...668L...3L} {668, L3}

\bibitem[\protect\citeauthoryear{{Lu}, {Minchev}, {Buck}, {Khoperskov},
  {Steinmetz}, {Libeskind}, {Cescutti}  \& {Freeman}}{{Lu} et~al.}{2022}]{lu22}
{Lu} Y.,  {Minchev} I.,  {Buck} T.,  {Khoperskov} S.,  {Steinmetz} M.,
  {Libeskind} N.,  {Cescutti} G.,   {Freeman} K.~C.,  2022, \mn@doi [arXiv
  e-prints] {10.48550/arXiv.2212.04515}, \href
  {https://ui.adsabs.harvard.edu/abs/2022arXiv221204515L} {p. arXiv:2212.04515}

\bibitem[\protect\citeauthoryear{{Ma}, {Hopkins}, {Wetzel}, {Kirby},
  {Angl{\'e}s-Alc{\'a}zar}, {Faucher-Gigu{\`e}re}, {Kere{\v{s}}}  \&
  {Quataert}}{{Ma} et~al.}{2017}]{2017MNRAS.467.2430M}
{Ma} X.,  {Hopkins} P.~F.,  {Wetzel} A.~R.,  {Kirby} E.~N.,
  {Angl{\'e}s-Alc{\'a}zar} D.,  {Faucher-Gigu{\`e}re} C.-A.,  {Kere{\v{s}}} D.,
    {Quataert} E.,  2017, \mn@doi [\mnras] {10.1093/mnras/stx273}, \href
  {https://ui.adsabs.harvard.edu/abs/2017MNRAS.467.2430M} {467, 2430}

\bibitem[\protect\citeauthoryear{{Marinacci}, {Pakmor}  \&
  {Springel}}{{Marinacci} et~al.}{2014}]{2014MNRAS.437.1750M}
{Marinacci} F.,  {Pakmor} R.,   {Springel} V.,  2014, \mn@doi [\mnras]
  {10.1093/mnras/stt2003}, \href
  {https://ui.adsabs.harvard.edu/abs/2014MNRAS.437.1750M} {437, 1750}

\bibitem[\protect\citeauthoryear{{Marinova} \& {Jogee}}{{Marinova} \&
  {Jogee}}{2007}]{2007ApJ...659.1176M}
{Marinova} I.,  {Jogee} S.,  2007, \mn@doi [\apj] {10.1086/512355}, \href
  {https://ui.adsabs.harvard.edu/abs/2007ApJ...659.1176M} {659, 1176}

\bibitem[\protect\citeauthoryear{{Martig}, {Minchev}  \& {Flynn}}{{Martig}
  et~al.}{2014}]{2014MNRAS.443.2452M}
{Martig} M.,  {Minchev} I.,   {Flynn} C.,  2014, \mn@doi [\mnras]
  {10.1093/mnras/stu1322}, \href
  {https://ui.adsabs.harvard.edu/abs/2014MNRAS.443.2452M} {443, 2452}

\bibitem[\protect\citeauthoryear{{Martinez-Valpuesta} \&
  {Shlosman}}{{Martinez-Valpuesta} \& {Shlosman}}{2004}]{2004ApJ...613L..29M}
{Martinez-Valpuesta} I.,  {Shlosman} I.,  2004, \mn@doi [\apjl]
  {10.1086/424876}, \href
  {https://ui.adsabs.harvard.edu/abs/2004ApJ...613L..29M} {613, L29}

\bibitem[\protect\citeauthoryear{{Martinez-Valpuesta}, {Shlosman}  \&
  {Heller}}{{Martinez-Valpuesta} et~al.}{2006}]{2006ApJ...637..214M}
{Martinez-Valpuesta} I.,  {Shlosman} I.,   {Heller} C.,  2006, \mn@doi [\apj]
  {10.1086/498338}, \href
  {https://ui.adsabs.harvard.edu/abs/2006ApJ...637..214M} {637, 214}

\bibitem[\protect\citeauthoryear{{Masters} et~al.,}{{Masters}
  et~al.}{2011}]{2011MNRAS.411.2026M}
{Masters} K.~L.,  et~al., 2011, \mn@doi [\mnras]
  {10.1111/j.1365-2966.2010.17834.x}, \href
  {https://ui.adsabs.harvard.edu/abs/2011MNRAS.411.2026M} {411, 2026}

\bibitem[\protect\citeauthoryear{{Masters} et~al.,}{{Masters}
  et~al.}{2012}]{2012MNRAS.424.2180M}
{Masters} K.~L.,  et~al., 2012, \mn@doi [\mnras]
  {10.1111/j.1365-2966.2012.21377.x}, \href
  {https://ui.adsabs.harvard.edu/abs/2012MNRAS.424.2180M} {424, 2180}

\bibitem[\protect\citeauthoryear{{Men{\'e}ndez-Delmestre}, {Sheth},
  {Schinnerer}, {Jarrett}  \& {Scoville}}{{Men{\'e}ndez-Delmestre}
  et~al.}{2007}]{2007ApJ...657..790M}
{Men{\'e}ndez-Delmestre} K.,  {Sheth} K.,  {Schinnerer} E.,  {Jarrett} T.~H.,
  {Scoville} N.~Z.,  2007, \mn@doi [\apj] {10.1086/511025}, \href
  {https://ui.adsabs.harvard.edu/abs/2007ApJ...657..790M} {657, 790}

\bibitem[\protect\citeauthoryear{{Minchev}, {Famaey}, {Quillen}, {Di Matteo},
  {Combes}, {Vlaji{\'c}}, {Erwin}  \& {Bland-Hawthorn}}{{Minchev}
  et~al.}{2012}]{minchev12a}
{Minchev} I.,  {Famaey} B.,  {Quillen} A.~C.,  {Di Matteo} P.,  {Combes} F.,
  {Vlaji{\'c}} M.,  {Erwin} P.,   {Bland-Hawthorn} J.,  2012, \mn@doi [\aap]
  {10.1051/0004-6361/201219198}, \href
  {https://ui.adsabs.harvard.edu/abs/2012A&A...548A.126M} {548, A126}

\bibitem[\protect\citeauthoryear{{Minchev}, {Chiappini}  \& {Martig}}{{Minchev}
  et~al.}{2013}]{minchev13}
{Minchev} I.,  {Chiappini} C.,   {Martig} M.,  2013, \mn@doi [\aap]
  {10.1051/0004-6361/201220189}, \href
  {https://ui.adsabs.harvard.edu/abs/2013A&A...558A...9M} {558, A9}

\bibitem[\protect\citeauthoryear{{Miwa} \& {Noguchi}}{{Miwa} \&
  {Noguchi}}{1998}]{1998ApJ...499..149M}
{Miwa} T.,  {Noguchi} M.,  1998, \mn@doi [\apj] {10.1086/305611}, \href
  {https://ui.adsabs.harvard.edu/abs/1998ApJ...499..149M} {499, 149}

\bibitem[\protect\citeauthoryear{{Moster}, {Naab}  \& {White}}{{Moster}
  et~al.}{2018}]{2018MNRAS.477.1822M}
{Moster} B.~P.,  {Naab} T.,   {White} S. D.~M.,  2018, \mn@doi [\mnras]
  {10.1093/mnras/sty655}, \href
  {https://ui.adsabs.harvard.edu/abs/2018MNRAS.477.1822M} {477, 1822}

\bibitem[\protect\citeauthoryear{{Motwani} et~al.,}{{Motwani}
  et~al.}{2022}]{2022ApJ...926..139M}
{Motwani} B.,  et~al., 2022, \mn@doi [\apj] {10.3847/1538-4357/ac3d2d}, \href
  {https://ui.adsabs.harvard.edu/abs/2022ApJ...926..139M} {926, 139}

\bibitem[\protect\citeauthoryear{{Nakada}, {Onaka}, {Yamamura}, {Deguchi},
  {Hashimoto}, {Izumiura}  \& {Sekiguchi}}{{Nakada}
  et~al.}{1991}]{1991Natur.353..140N}
{Nakada} Y.,  {Onaka} T.,  {Yamamura} I.,  {Deguchi} S.,  {Hashimoto} O.,
  {Izumiura} H.,   {Sekiguchi} K.,  1991, \mn@doi [\nat] {10.1038/353140a0},
  \href {https://ui.adsabs.harvard.edu/abs/1991Natur.353..140N} {353, 140}

\bibitem[\protect\citeauthoryear{{Neeleman}, {Prochaska}, {Kanekar}  \&
  {Rafelski}}{{Neeleman} et~al.}{2020}]{2020Natur.581..269N}
{Neeleman} M.,  {Prochaska} J.~X.,  {Kanekar} N.,   {Rafelski} M.,  2020,
  \mn@doi [\nat] {10.1038/s41586-020-2276-y}, \href
  {https://ui.adsabs.harvard.edu/abs/2020Natur.581..269N} {581, 269}

\bibitem[\protect\citeauthoryear{{Nelson} et~al.,}{{Nelson}
  et~al.}{2019a}]{2019ComAC...6....2N}
{Nelson} D.,  et~al., 2019a, \mn@doi [Computational Astrophysics and Cosmology]
  {10.1186/s40668-019-0028-x}, \href
  {https://ui.adsabs.harvard.edu/abs/2019ComAC...6....2N} {6, 2}

\bibitem[\protect\citeauthoryear{{Nelson} et~al.,}{{Nelson}
  et~al.}{2019b}]{2019MNRAS.490.3234N}
{Nelson} D.,  et~al., 2019b, \mn@doi [\mnras] {10.1093/mnras/stz2306}, \href
  {https://ui.adsabs.harvard.edu/abs/2019MNRAS.490.3234N} {490, 3234}

\bibitem[\protect\citeauthoryear{{Nelson} et~al.,}{{Nelson}
  et~al.}{2020}]{2020MNRAS.498.2391N}
{Nelson} D.,  et~al., 2020, \mn@doi [\mnras] {10.1093/mnras/staa2419}, \href
  {https://ui.adsabs.harvard.edu/abs/2020MNRAS.498.2391N} {498, 2391}

\bibitem[\protect\citeauthoryear{{Nelson} et~al.,}{{Nelson}
  et~al.}{2021}]{2021MNRAS.508..219N}
{Nelson} E.~J.,  et~al., 2021, \mn@doi [\mnras] {10.1093/mnras/stab2131}, \href
  {https://ui.adsabs.harvard.edu/abs/2021MNRAS.508..219N} {508, 219}

\bibitem[\protect\citeauthoryear{{Nepal} et~al.,}{{Nepal}
  et~al.}{2024}]{2024A&A...681L...8N}
{Nepal} S.,  et~al., 2024, \mn@doi [\aap] {10.1051/0004-6361/202348365}, \href
  {https://ui.adsabs.harvard.edu/abs/2024A&A...681L...8N} {681, L8}

\bibitem[\protect\citeauthoryear{{Nidever} et~al.,}{{Nidever}
  et~al.}{2014}]{2014ApJ...796...38N}
{Nidever} D.~L.,  et~al., 2014, \mn@doi [\apj] {10.1088/0004-637X/796/1/38},
  \href {https://ui.adsabs.harvard.edu/abs/2014ApJ...796...38N} {796, 38}

\bibitem[\protect\citeauthoryear{{Nogueras-Lara} et~al.,}{{Nogueras-Lara}
  et~al.}{2020}]{2020NatAs...4..377N}
{Nogueras-Lara} F.,  et~al., 2020, \mn@doi [Nature Astronomy]
  {10.1038/s41550-019-0967-9}, \href
  {https://ui.adsabs.harvard.edu/abs/2020NatAs...4..377N} {4, 377}

\bibitem[\protect\citeauthoryear{{Nogueras-Lara}, {Schultheis}, {Najarro},
  {Sormani}, {Gadotti}  \& {Rich}}{{Nogueras-Lara}
  et~al.}{2023}]{2023A&A...671L..10N}
{Nogueras-Lara} F.,  {Schultheis} M.,  {Najarro} F.,  {Sormani} M.~C.,
  {Gadotti} D.~A.,   {Rich} R.~M.,  2023, \mn@doi [\aap]
  {10.1051/0004-6361/202345941}, \href
  {https://ui.adsabs.harvard.edu/abs/2023A&A...671L..10N} {671, L10}

\bibitem[\protect\citeauthoryear{{Ostriker} \& {Peebles}}{{Ostriker} \&
  {Peebles}}{1973}]{1973ApJ...186..467O}
{Ostriker} J.~P.,  {Peebles} P.~J.~E.,  1973, \mn@doi [\apj] {10.1086/152513},
  \href {https://ui.adsabs.harvard.edu/abs/1973ApJ...186..467O} {186, 467}

\bibitem[\protect\citeauthoryear{{Pahwa} et~al.,}{{Pahwa}
  et~al.}{2016}]{2016MNRAS.457..695P}
{Pahwa} I.,  et~al., 2016, \mn@doi [\mnras] {10.1093/mnras/stv2930}, \href
  {https://ui.adsabs.harvard.edu/abs/2016MNRAS.457..695P} {457, 695}

\bibitem[\protect\citeauthoryear{{Parry}, {Eke}  \& {Frenk}}{{Parry}
  et~al.}{2009}]{2009MNRAS.396.1972P}
{Parry} O.~H.,  {Eke} V.~R.,   {Frenk} C.~S.,  2009, \mn@doi [\mnras]
  {10.1111/j.1365-2966.2009.14921.x}, \href
  {https://ui.adsabs.harvard.edu/abs/2009MNRAS.396.1972P} {396, 1972}

\bibitem[\protect\citeauthoryear{{Peschken} \& {{\L}okas}}{{Peschken} \&
  {{\L}okas}}{2019}]{2019MNRAS.483.2721P}
{Peschken} N.,  {{\L}okas} E.~L.,  2019, \mn@doi [\mnras]
  {10.1093/mnras/sty3277}, \href
  {https://ui.adsabs.harvard.edu/abs/2019MNRAS.483.2721P} {483, 2721}

\bibitem[\protect\citeauthoryear{{Pillepich} et~al.,}{{Pillepich}
  et~al.}{2018}]{2018MNRAS.473.4077P}
{Pillepich} A.,  et~al., 2018, \mn@doi [\mnras] {10.1093/mnras/stx2656}, \href
  {https://ui.adsabs.harvard.edu/abs/2018MNRAS.473.4077P} {473, 4077}

\bibitem[\protect\citeauthoryear{{Pillepich} et~al.,}{{Pillepich}
  et~al.}{2019}]{2019MNRAS.490.3196P}
{Pillepich} A.,  et~al., 2019, \mn@doi [\mnras] {10.1093/mnras/stz2338}, \href
  {https://ui.adsabs.harvard.edu/abs/2019MNRAS.490.3196P} {490, 3196}

\bibitem[\protect\citeauthoryear{{Pillepich} et~al.,}{{Pillepich}
  et~al.}{2023}]{2023arXiv230316217P}
{Pillepich} A.,  et~al., 2023, \mn@doi [arXiv e-prints]
  {10.48550/arXiv.2303.16217}, \href
  {https://ui.adsabs.harvard.edu/abs/2023arXiv230316217P} {p. arXiv:2303.16217}

\bibitem[\protect\citeauthoryear{{Polyachenko}, {Berczik}  \&
  {Just}}{{Polyachenko} et~al.}{2016}]{2016MNRAS.462.3727P}
{Polyachenko} E.~V.,  {Berczik} P.,   {Just} A.,  2016, \mn@doi [\mnras]
  {10.1093/mnras/stw1907}, \href
  {https://ui.adsabs.harvard.edu/abs/2016MNRAS.462.3727P} {462, 3727}

\bibitem[\protect\citeauthoryear{{Posti} \& {Helmi}}{{Posti} \&
  {Helmi}}{2019}]{2019A&A...621A..56P}
{Posti} L.,  {Helmi} A.,  2019, \mn@doi [\aap] {10.1051/0004-6361/201833355},
  \href {https://ui.adsabs.harvard.edu/abs/2019A&A...621A..56P} {621, A56}

\bibitem[\protect\citeauthoryear{{Prendergast}}{{Prendergast}}{1983}]{1983IAUS..100..215P}
{Prendergast} K.~H.,  1983, in {Athanassoula} E.,  ed., ~ Vol. 100, Internal
  Kinematics and Dynamics of Galaxies. pp 215--220

\bibitem[\protect\citeauthoryear{{Queiroz} et~al.,}{{Queiroz}
  et~al.}{2021}]{2021A&A...656A.156Q}
{Queiroz} A.~B.~A.,  et~al., 2021, \mn@doi [\aap]
  {10.1051/0004-6361/202039030}, \href
  {https://ui.adsabs.harvard.edu/abs/2021A&A...656A.156Q} {656, A156}

\bibitem[\protect\citeauthoryear{{Ratcliffe} et~al.,}{{Ratcliffe}
  et~al.}{2023}]{ratcliffe23}
{Ratcliffe} B.,  et~al., 2023, \mn@doi [\mnras] {10.1093/mnras/stad1573}, \href
  {https://ui.adsabs.harvard.edu/abs/2023MNRAS.525.2208R} {525, 2208}

\bibitem[\protect\citeauthoryear{{Reddish} et~al.,}{{Reddish}
  et~al.}{2022}]{2022MNRAS.512..160R}
{Reddish} J.,  et~al., 2022, \mn@doi [\mnras] {10.1093/mnras/stac494}, \href
  {https://ui.adsabs.harvard.edu/abs/2022MNRAS.512..160R} {512, 160}

\bibitem[\protect\citeauthoryear{{Renaud} et~al.,}{{Renaud}
  et~al.}{2015}]{2015MNRAS.454.3299R}
{Renaud} F.,  et~al., 2015, \mn@doi [\mnras] {10.1093/mnras/stv2223}, \href
  {https://ui.adsabs.harvard.edu/abs/2015MNRAS.454.3299R} {454, 3299}

\bibitem[\protect\citeauthoryear{{Renaud}, {Agertz}, {Read}, {Ryde},
  {Andersson}, {Bensby}, {Rey}  \& {Feuillet}}{{Renaud}
  et~al.}{2021a}]{2021MNRAS.503.5846R}
{Renaud} F.,  {Agertz} O.,  {Read} J.~I.,  {Ryde} N.,  {Andersson} E.~P.,
  {Bensby} T.,  {Rey} M.~P.,   {Feuillet} D.~K.,  2021a, \mn@doi [\mnras]
  {10.1093/mnras/stab250}, \href
  {https://ui.adsabs.harvard.edu/abs/2021MNRAS.503.5846R} {503, 5846}

\bibitem[\protect\citeauthoryear{{Renaud}, {Agertz}, {Andersson}, {Read},
  {Ryde}, {Bensby}, {Rey}  \& {Feuillet}}{{Renaud}
  et~al.}{2021b}]{2021MNRAS.503.5868R}
{Renaud} F.,  {Agertz} O.,  {Andersson} E.~P.,  {Read} J.~I.,  {Ryde} N.,
  {Bensby} T.,  {Rey} M.~P.,   {Feuillet} D.~K.,  2021b, \mn@doi [\mnras]
  {10.1093/mnras/stab543}, \href
  {https://ui.adsabs.harvard.edu/abs/2021MNRAS.503.5868R} {503, 5868}

\bibitem[\protect\citeauthoryear{{Renaud}, {Romeo}  \& {Agertz}}{{Renaud}
  et~al.}{2021c}]{2021MNRAS.508..352R}
{Renaud} F.,  {Romeo} A.~B.,   {Agertz} O.,  2021c, \mn@doi [\mnras]
  {10.1093/mnras/stab2604}, \href
  {https://ui.adsabs.harvard.edu/abs/2021MNRAS.508..352R} {508, 352}

\bibitem[\protect\citeauthoryear{{Rix} et~al.,}{{Rix}
  et~al.}{2022}]{2022ApJ...941...45R}
{Rix} H.-W.,  et~al., 2022, \mn@doi [\apj] {10.3847/1538-4357/ac9e01}, \href
  {https://ui.adsabs.harvard.edu/abs/2022ApJ...941...45R} {941, 45}

\bibitem[\protect\citeauthoryear{{Rizzo}, {Vegetti}, {Powell}, {Fraternali},
  {McKean}, {Stacey}  \& {White}}{{Rizzo} et~al.}{2020}]{2020Natur.584..201R}
{Rizzo} F.,  {Vegetti} S.,  {Powell} D.,  {Fraternali} F.,  {McKean} J.~P.,
  {Stacey} H.~R.,   {White} S.~D.~M.,  2020, \mn@doi [\nat]
  {10.1038/s41586-020-2572-6}, \href
  {https://ui.adsabs.harvard.edu/abs/2020Natur.584..201R} {584, 201}

\bibitem[\protect\citeauthoryear{{Romanowsky} \& {Fall}}{{Romanowsky} \&
  {Fall}}{2012}]{2012ApJS..203...17R}
{Romanowsky} A.~J.,  {Fall} S.~M.,  2012, \mn@doi [\apjs]
  {10.1088/0067-0049/203/2/17}, \href
  {https://ui.adsabs.harvard.edu/abs/2012ApJS..203...17R} {203, 17}

\bibitem[\protect\citeauthoryear{{Romeo}, {Agertz}  \& {Renaud}}{{Romeo}
  et~al.}{2023}]{2023MNRAS.518.1002R}
{Romeo} A.~B.,  {Agertz} O.,   {Renaud} F.,  2023, \mn@doi [\mnras]
  {10.1093/mnras/stac3074}, \href
  {https://ui.adsabs.harvard.edu/abs/2023MNRAS.518.1002R} {518, 1002}

\bibitem[\protect\citeauthoryear{{Rosas-Guevara} et~al.,}{{Rosas-Guevara}
  et~al.}{2022}]{2022MNRAS.512.5339R}
{Rosas-Guevara} Y.,  et~al., 2022, \mn@doi [\mnras] {10.1093/mnras/stac816},
  \href {https://ui.adsabs.harvard.edu/abs/2022MNRAS.512.5339R} {512, 5339}

\bibitem[\protect\citeauthoryear{{Sales}, {Navarro}, {Theuns}, {Schaye},
  {White}, {Frenk}, {Crain}  \& {Dalla Vecchia}}{{Sales}
  et~al.}{2012}]{2012MNRAS.423.1544S}
{Sales} L.~V.,  {Navarro} J.~F.,  {Theuns} T.,  {Schaye} J.,  {White} S. D.~M.,
   {Frenk} C.~S.,  {Crain} R.~A.,   {Dalla Vecchia} C.,  2012, \mn@doi [\mnras]
  {10.1111/j.1365-2966.2012.20975.x}, \href
  {https://ui.adsabs.harvard.edu/abs/2012MNRAS.423.1544S} {423, 1544}

\bibitem[\protect\citeauthoryear{{Sandage}}{{Sandage}}{1961}]{1961hag..book.....S}
{Sandage} A.,  1961, {The Hubble Atlas of Galaxies}

\bibitem[\protect\citeauthoryear{{Sanders}, {Kawata}, {Matsunaga}, {Sormani},
  {Smith}, {Minniti}  \& {Gerhard}}{{Sanders}
  et~al.}{2024}]{2024MNRAS.tmp..735S}
{Sanders} J.~L.,  {Kawata} D.,  {Matsunaga} N.,  {Sormani} M.~C.,  {Smith}
  L.~C.,  {Minniti} D.,   {Gerhard} O.,  2024, \mn@doi [\mnras]
  {10.1093/mnras/stae711}, \href
  {https://ui.adsabs.harvard.edu/abs/2024MNRAS.tmp..735S} {}

\bibitem[\protect\citeauthoryear{{Sch{\"o}del} et~al.,}{{Sch{\"o}del}
  et~al.}{2023}]{2023A&A...672L...8S}
{Sch{\"o}del} R.,  et~al., 2023, \mn@doi [\aap] {10.1051/0004-6361/202346335},
  \href {https://ui.adsabs.harvard.edu/abs/2023A&A...672L...8S} {672, L8}

\bibitem[\protect\citeauthoryear{{Segovia Otero}, {Renaud}  \&
  {Agertz}}{{Segovia Otero} et~al.}{2022}]{2022MNRAS.516.2272S}
{Segovia Otero} {\'A}.,  {Renaud} F.,   {Agertz} O.,  2022, \mn@doi [\mnras]
  {10.1093/mnras/stac2368}, \href
  {https://ui.adsabs.harvard.edu/abs/2022MNRAS.516.2272S} {516, 2272}

\bibitem[\protect\citeauthoryear{{Sellwood}}{{Sellwood}}{1980}]{1980A&A....89..296S}
{Sellwood} J.~A.,  1980, \aap, \href
  {https://ui.adsabs.harvard.edu/abs/1980A&A....89..296S} {89, 296}

\bibitem[\protect\citeauthoryear{{Sellwood} \& {Carlberg}}{{Sellwood} \&
  {Carlberg}}{2023}]{2023arXiv230214775S}
{Sellwood} J.~A.,  {Carlberg} R.~G.,  2023, \mn@doi [arXiv e-prints]
  {10.48550/arXiv.2302.14775}, \href
  {https://ui.adsabs.harvard.edu/abs/2023arXiv230214775S} {p. arXiv:2302.14775}

\bibitem[\protect\citeauthoryear{{Sellwood} \& {Wilkinson}}{{Sellwood} \&
  {Wilkinson}}{1993}]{1993RPPh...56..173S}
{Sellwood} J.~A.,  {Wilkinson} A.,  1993, \mn@doi [Reports on Progress in
  Physics] {10.1088/0034-4885/56/2/001}, \href
  {https://ui.adsabs.harvard.edu/abs/1993RPPh...56..173S} {56, 173}

\bibitem[\protect\citeauthoryear{{Semenov}, {Conroy}, {Chandra}, {Hernquist}
  \& {Nelson}}{{Semenov} et~al.}{2023a}]{2023arXiv230609398S}
{Semenov} V.~A.,  {Conroy} C.,  {Chandra} V.,  {Hernquist} L.,   {Nelson} D.,
  2023a, \mn@doi [arXiv e-prints] {10.48550/arXiv.2306.09398}, \href
  {https://ui.adsabs.harvard.edu/abs/2023arXiv230609398S} {p. arXiv:2306.09398}

\bibitem[\protect\citeauthoryear{{Semenov}, {Conroy}, {Chandra}, {Hernquist}
  \& {Nelson}}{{Semenov} et~al.}{2023b}]{2023arXiv230613125S}
{Semenov} V.~A.,  {Conroy} C.,  {Chandra} V.,  {Hernquist} L.,   {Nelson} D.,
  2023b, \mn@doi [arXiv e-prints] {10.48550/arXiv.2306.13125}, \href
  {https://ui.adsabs.harvard.edu/abs/2023arXiv230613125S} {p. arXiv:2306.13125}

\bibitem[\protect\citeauthoryear{{Sestito} et~al.,}{{Sestito}
  et~al.}{2020}]{2020MNRAS.497L...7S}
{Sestito} F.,  et~al., 2020, \mn@doi [\mnras] {10.1093/mnrasl/slaa022}, \href
  {https://ui.adsabs.harvard.edu/abs/2020MNRAS.497L...7S} {497, L7}

\bibitem[\protect\citeauthoryear{{Sheth} et~al.,}{{Sheth}
  et~al.}{2008}]{2008ApJ...675.1141S}
{Sheth} K.,  et~al., 2008, \mn@doi [\apj] {10.1086/524980}, \href
  {https://ui.adsabs.harvard.edu/abs/2008ApJ...675.1141S} {675, 1141}

\bibitem[\protect\citeauthoryear{{Simons} et~al.,}{{Simons}
  et~al.}{2017}]{2017ApJ...843...46S}
{Simons} R.~C.,  et~al., 2017, \mn@doi [\apj] {10.3847/1538-4357/aa740c}, \href
  {https://ui.adsabs.harvard.edu/abs/2017ApJ...843...46S} {843, 46}

\bibitem[\protect\citeauthoryear{{Snaith}, {Haywood}, {Di Matteo}, {Lehnert},
  {Combes}, {Katz}  \& {G{\'o}mez}}{{Snaith}
  et~al.}{2015}]{2015A&A...578A..87S}
{Snaith} O.,  {Haywood} M.,  {Di Matteo} P.,  {Lehnert} M.~D.,  {Combes} F.,
  {Katz} D.,   {G{\'o}mez} A.,  2015, \mn@doi [\aap]
  {10.1051/0004-6361/201424281}, \href
  {https://ui.adsabs.harvard.edu/abs/2015A&A...578A..87S} {578, A87}

\bibitem[\protect\citeauthoryear{{Sotillo-Ramos} et~al.,}{{Sotillo-Ramos}
  et~al.}{2022}]{2022MNRAS.516.5404S}
{Sotillo-Ramos} D.,  et~al., 2022, \mn@doi [\mnras] {10.1093/mnras/stac2586},
  \href {https://ui.adsabs.harvard.edu/abs/2022MNRAS.516.5404S} {516, 5404}

\bibitem[\protect\citeauthoryear{{Springel}}{{Springel}}{2010}]{2010MNRAS.401..791S}
{Springel} V.,  2010, \mn@doi [\mnras] {10.1111/j.1365-2966.2009.15715.x},
  \href {https://ui.adsabs.harvard.edu/abs/2010MNRAS.401..791S} {401, 791}

\bibitem[\protect\citeauthoryear{{Stanek}, {Udalski}, {Szyma{\'N}ski},
  {Ka{\L}u{\.Z}ny}, {Kubiak}, {Mateo}  \& {Krzemi{\'N}ski}}{{Stanek}
  et~al.}{1997}]{1997ApJ...477..163S}
{Stanek} K.~Z.,  {Udalski} A.,  {Szyma{\'N}ski} M.,  {Ka{\L}u{\.Z}ny} J.,
  {Kubiak} Z.~M.,  {Mateo} M.,   {Krzemi{\'N}ski} W.,  1997, \mn@doi [\apj]
  {10.1086/303702}, \href
  {https://ui.adsabs.harvard.edu/abs/1997ApJ...477..163S} {477, 163}

\bibitem[\protect\citeauthoryear{{Steinmetz} \& {Navarro}}{{Steinmetz} \&
  {Navarro}}{2002}]{2002NewA....7..155S}
{Steinmetz} M.,  {Navarro} J.~F.,  2002, \mn@doi [\na]
  {10.1016/S1384-1076(02)00102-1}, \href
  {https://ui.adsabs.harvard.edu/abs/2002NewA....7..155S} {7, 155}

\bibitem[\protect\citeauthoryear{{Stewart}, {Brooks}, {Bullock}, {Maller},
  {Diemand}, {Wadsley}  \& {Moustakas}}{{Stewart}
  et~al.}{2013}]{2013ApJ...769...74S}
{Stewart} K.~R.,  {Brooks} A.~M.,  {Bullock} J.~S.,  {Maller} A.~H.,  {Diemand}
  J.,  {Wadsley} J.,   {Moustakas} L.~A.,  2013, \mn@doi [\apj]
  {10.1088/0004-637X/769/1/74}, \href
  {https://ui.adsabs.harvard.edu/abs/2013ApJ...769...74S} {769, 74}

\bibitem[\protect\citeauthoryear{{Teklu}, {Remus}, {Dolag}, {Beck}, {Burkert},
  {Schmidt}, {Schulze}  \& {Steinborn}}{{Teklu}
  et~al.}{2015}]{2015ApJ...812...29T}
{Teklu} A.~F.,  {Remus} R.-S.,  {Dolag} K.,  {Beck} A.~M.,  {Burkert} A.,
  {Schmidt} A.~S.,  {Schulze} F.,   {Steinborn} L.~K.,  2015, \mn@doi [\apj]
  {10.1088/0004-637X/812/1/29}, \href
  {https://ui.adsabs.harvard.edu/abs/2015ApJ...812...29T} {812, 29}

\bibitem[\protect\citeauthoryear{{Tempel} \& {Libeskind}}{{Tempel} \&
  {Libeskind}}{2013}]{2013ApJ...775L..42T}
{Tempel} E.,  {Libeskind} N.~I.,  2013, \mn@doi [\apjl]
  {10.1088/2041-8205/775/2/L42}, \href
  {https://ui.adsabs.harvard.edu/abs/2013ApJ...775L..42T} {775, L42}

\bibitem[\protect\citeauthoryear{{Thorsbro} et~al.,}{{Thorsbro}
  et~al.}{2020}]{2020ApJ...894...26T}
{Thorsbro} B.,  et~al., 2020, \mn@doi [\apj] {10.3847/1538-4357/ab8226}, \href
  {https://ui.adsabs.harvard.edu/abs/2020ApJ...894...26T} {894, 26}

\bibitem[\protect\citeauthoryear{{Tissera}, {Machado}, {Carollo}, {Minniti},
  {Beers}, {Zoccali}  \& {Meza}}{{Tissera} et~al.}{2018}]{2018MNRAS.473.1656T}
{Tissera} P.~B.,  {Machado} R. E.~G.,  {Carollo} D.,  {Minniti} D.,  {Beers}
  T.~C.,  {Zoccali} M.,   {Meza} A.,  2018, \mn@doi [\mnras]
  {10.1093/mnras/stx2431}, \href
  {https://ui.adsabs.harvard.edu/abs/2018MNRAS.473.1656T} {473, 1656}

\bibitem[\protect\citeauthoryear{{Toomre}}{{Toomre}}{1964}]{1964ApJ...139.1217T}
{Toomre} A.,  1964, \mn@doi [\apj] {10.1086/147861}, \href
  {https://ui.adsabs.harvard.edu/abs/1964ApJ...139.1217T} {139, 1217}

\bibitem[\protect\citeauthoryear{{Toomre}}{{Toomre}}{1981}]{1981seng.proc..111T}
{Toomre} A.,  1981, in {Fall} S.~M.,  {Lynden-Bell} D.,  eds, Structure and
  Evolution of Normal Galaxies. pp 111--136

\bibitem[\protect\citeauthoryear{{Vasiliev}, {Belokurov}  \&
  {Erkal}}{{Vasiliev} et~al.}{2021}]{2021MNRAS.501.2279V}
{Vasiliev} E.,  {Belokurov} V.,   {Erkal} D.,  2021, \mn@doi [\mnras]
  {10.1093/mnras/staa3673}, \href
  {https://ui.adsabs.harvard.edu/abs/2021MNRAS.501.2279V} {501, 2279}

\bibitem[\protect\citeauthoryear{{Vislosky} et~al.,}{{Vislosky}
  et~al.}{2024}]{2024MNRAS.528.3576V}
{Vislosky} E.,  et~al., 2024, \mn@doi [\mnras] {10.1093/mnras/stae083}, \href
  {https://ui.adsabs.harvard.edu/abs/2024MNRAS.528.3576V} {528, 3576}

\bibitem[\protect\citeauthoryear{{Vogelsberger}, {Genel}, {Sijacki}, {Torrey},
  {Springel}  \& {Hernquist}}{{Vogelsberger}
  et~al.}{2013}]{2013MNRAS.436.3031V}
{Vogelsberger} M.,  {Genel} S.,  {Sijacki} D.,  {Torrey} P.,  {Springel} V.,
  {Hernquist} L.,  2013, \mn@doi [\mnras] {10.1093/mnras/stt1789}, \href
  {https://ui.adsabs.harvard.edu/abs/2013MNRAS.436.3031V} {436, 3031}

\bibitem[\protect\citeauthoryear{{Vogelsberger} et~al.,}{{Vogelsberger}
  et~al.}{2014}]{2014Natur.509..177V}
{Vogelsberger} M.,  et~al., 2014, \mn@doi [\nat] {10.1038/nature13316}, \href
  {https://ui.adsabs.harvard.edu/abs/2014Natur.509..177V} {509, 177}

\bibitem[\protect\citeauthoryear{{Wang} et~al.,}{{Wang}
  et~al.}{2012}]{2012MNRAS.423.3486W}
{Wang} J.,  et~al., 2012, \mn@doi [\mnras] {10.1111/j.1365-2966.2012.21147.x},
  \href {https://ui.adsabs.harvard.edu/abs/2012MNRAS.423.3486W} {423, 3486}

\bibitem[\protect\citeauthoryear{{Wang}, {Guo}, {Kang}  \& {Libeskind}}{{Wang}
  et~al.}{2018}]{2018ApJ...866..138W}
{Wang} P.,  {Guo} Q.,  {Kang} X.,   {Libeskind} N.~I.,  2018, \mn@doi [\apj]
  {10.3847/1538-4357/aae20f}, \href
  {https://ui.adsabs.harvard.edu/abs/2018ApJ...866..138W} {866, 138}

\bibitem[\protect\citeauthoryear{{Wang}, {Libeskind}, {Tempel}, {Kang}  \&
  {Guo}}{{Wang} et~al.}{2021}]{2021NatAs...5..839W}
{Wang} P.,  {Libeskind} N.~I.,  {Tempel} E.,  {Kang} X.,   {Guo} Q.,  2021,
  \mn@doi [Nature Astronomy] {10.1038/s41550-021-01380-6}, \href
  {https://ui.adsabs.harvard.edu/abs/2021NatAs...5..839W} {5, 839}

\bibitem[\protect\citeauthoryear{{Wegg} \& {Gerhard}}{{Wegg} \&
  {Gerhard}}{2013}]{2013MNRAS.435.1874W}
{Wegg} C.,  {Gerhard} O.,  2013, \mn@doi [\mnras] {10.1093/mnras/stt1376},
  \href {https://ui.adsabs.harvard.edu/abs/2013MNRAS.435.1874W} {435, 1874}

\bibitem[\protect\citeauthoryear{{Weiland} et~al.,}{{Weiland}
  et~al.}{1994}]{1994ApJ...425L..81W}
{Weiland} J.~L.,  et~al., 1994, \mn@doi [\apjl] {10.1086/187315}, \href
  {https://ui.adsabs.harvard.edu/abs/1994ApJ...425L..81W} {425, L81}

\bibitem[\protect\citeauthoryear{{Weinberger} et~al.,}{{Weinberger}
  et~al.}{2017}]{2017MNRAS.465.3291W}
{Weinberger} R.,  et~al., 2017, \mn@doi [\mnras] {10.1093/mnras/stw2944}, \href
  {https://ui.adsabs.harvard.edu/abs/2017MNRAS.465.3291W} {465, 3291}

\bibitem[\protect\citeauthoryear{{Weinberger}, {Springel}  \&
  {Pakmor}}{{Weinberger} et~al.}{2020}]{2020ApJS..248...32W}
{Weinberger} R.,  {Springel} V.,   {Pakmor} R.,  2020, \mn@doi [\apjs]
  {10.3847/1538-4365/ab908c}, \href
  {https://ui.adsabs.harvard.edu/abs/2020ApJS..248...32W} {248, 32}

\bibitem[\protect\citeauthoryear{{Yu} et~al.,}{{Yu}
  et~al.}{2021}]{2021MNRAS.505..889Y}
{Yu} S.,  et~al., 2021, \mn@doi [\mnras] {10.1093/mnras/stab1339}, \href
  {https://ui.adsabs.harvard.edu/abs/2021MNRAS.505..889Y} {505, 889}

\bibitem[\protect\citeauthoryear{{Zanisi} et~al.,}{{Zanisi}
  et~al.}{2021}]{2021MNRAS.501.4359Z}
{Zanisi} L.,  et~al., 2021, \mn@doi [\mnras] {10.1093/mnras/staa3864}, \href
  {https://ui.adsabs.harvard.edu/abs/2021MNRAS.501.4359Z} {501, 4359}

\bibitem[\protect\citeauthoryear{{Zhao}, {Du}, {Ho}, {Debattista}  \&
  {Shi}}{{Zhao} et~al.}{2020}]{2020ApJ...904..170Z}
{Zhao} D.,  {Du} M.,  {Ho} L.~C.,  {Debattista} V.~P.,   {Shi} J.,  2020,
  \mn@doi [\apj] {10.3847/1538-4357/abbe1b}, \href
  {https://ui.adsabs.harvard.edu/abs/2020ApJ...904..170Z} {904, 170}

\bibitem[\protect\citeauthoryear{{Zhou}, {Zhu}, {Wang}  \& {Feng}}{{Zhou}
  et~al.}{2020}]{2020ApJ...895...92Z}
{Zhou} Z.-B.,  {Zhu} W.,  {Wang} Y.,   {Feng} L.-L.,  2020, \mn@doi [\apj]
  {10.3847/1538-4357/ab8d32}, \href
  {https://ui.adsabs.harvard.edu/abs/2020ApJ...895...92Z} {895, 92}

\bibitem[\protect\citeauthoryear{{de S{\'a}-Freitas} et~al.,}{{de
  S{\'a}-Freitas} et~al.}{2023}]{2023A&A...671A...8D}
{de S{\'a}-Freitas} C.,  et~al., 2023, \mn@doi [\aap]
  {10.1051/0004-6361/202244667}, \href
  {https://ui.adsabs.harvard.edu/abs/2023A&A...671A...8D} {671, A8}

\bibitem[\protect\citeauthoryear{{de Vaucouleurs}}{{de
  Vaucouleurs}}{1959}]{1959HDP....53..275D}
{de Vaucouleurs} G.,  1959, \mn@doi [Handbuch der Physik]
  {10.1007/978-3-642-45932-0_7}, \href
  {https://ui.adsabs.harvard.edu/abs/1959HDP....53..275D} {53, 275}

\bibitem[\protect\citeauthoryear{{van Donkelaar}, {Agertz}  \& {Renaud}}{{van
  Donkelaar} et~al.}{2022}]{2022MNRAS.512.3806V}
{van Donkelaar} F.,  {Agertz} O.,   {Renaud} F.,  2022, \mn@doi [\mnras]
  {10.1093/mnras/stac692}, \href
  {https://ui.adsabs.harvard.edu/abs/2022MNRAS.512.3806V} {512, 3806}

\makeatother
\end{thebibliography}

\bsp	
\label{lastpage}
\end{document}